\documentclass[10pt,journal]{IEEEtran}
\ifCLASSINFOpdf

\else

\fi

\usepackage{mathrsfs}
\usepackage{amsmath}
\usepackage{amsfonts}
\usepackage{amsthm}
\usepackage{amssymb}
\usepackage{graphicx}
\usepackage{subfigure}
\usepackage{indentfirst}
\usepackage{array}
\usepackage{epstopdf}
\usepackage{cite}
\usepackage{enumerate}
\usepackage{bm}
\usepackage{dsfont}
\usepackage{multirow}
\usepackage{verbatim}
\usepackage{stfloats}
\usepackage{makecell}
\usepackage{cancel}
\usepackage{threeparttable}
\usepackage{tikz}
\usetikzlibrary{positioning}
\usetikzlibrary{spy}
\usepackage{wrapfig}
\usepackage{pgfplots}
\usepackage{nicefrac}
\usepackage{thmtools,thm-restate}
\usepackage{color}

\usepackage{booktabs}
\usepackage{graphicx}
\usepackage{bbm}
\usepackage{adjustbox}
\usepackage{multirow}
\usepackage{tabularx}
\usepackage[misc]{ifsym}
\usepackage{bbding}
\usepackage{colortbl}

\newcolumntype{Y}{>{\centering\arraybackslash}X}
\newcolumntype{C}[1]{>{\centering\arraybackslash}m{#1}}
\usepackage{algorithm}
\usepackage{algorithmicx}
\usepackage{algpseudocode}
\usepackage{xcolor}
\usepackage[linkcolor=blue,citecolor=blue,colorlinks=true,urlcolor=blue]{hyperref}

\algnewcommand{\LineComment}[1]{\State {\color{gray}\(\triangleright\) #1}}
\definecolor{lightblue}{rgb}{0.9, 0.92, 0.96}
\definecolor{subsectioncolor}{rgb}{0, 0, 0.706}   
\definecolor{mylightblue}{rgb}{0.851, 0.898, 0.941} %
\definecolor{LightCyan}{rgb}{0.88,1,1}
\definecolor{LightOrange}{rgb}{1,0.85,0.70}

\usepackage{pifont}
\definecolor{C0}{rgb}{0.121569, 0.466667, 0.705882}
\definecolor{C1}{rgb}{1.000000, 0.498039, 0.054902}
\definecolor{C2}{rgb}{0.172549, 0.627451, 0.172549}
\definecolor{C3}{rgb}{0.839216, 0.152941, 0.156863}
\definecolor{C4}{rgb}{0.580392, 0.403922, 0.741176}
\definecolor{C5}{rgb}{0.549020, 0.337255, 0.294118}
\definecolor{C6}{rgb}{0.890196, 0.466667, 0.760784}
\definecolor{C7}{rgb}{0.498039, 0.498039, 0.498039}
\definecolor{C8}{rgb}{0.737255, 0.741176, 0.133333}
\definecolor{C9}{rgb}{0.090196, 0.745098, 0.811765}
\definecolor{trolleygrey}{rgb}{0.5, 0.5, 0.5}

\newcommand{\add}[1] {#1} % for non-tracked version
\usepackage{enumitem}

\begin{document}
	
	\title{DiT-JSCC: Rethinking Deep JSCC with Diffusion Transformers and Semantic Representations}

	\author{Kailin Tan,
		Jincheng Dai,~\IEEEmembership{Member, IEEE},
		Sixian Wang,~\IEEEmembership{Member,~IEEE},
		Guo Lu,~\IEEEmembership{Member,~IEEE},
		Shuo Shao,~\IEEEmembership{Member,~IEEE},
		Kai Niu,~\IEEEmembership{Member,~IEEE},
		Wenjun Zhang,~\IEEEmembership{Fellow, IEEE},
		and Ping Zhang,~\IEEEmembership{Fellow, IEEE}
		
		\thanks{This work was supported in part by the National Key Research and Development Program of China under Grant 2024YFF0509700, in part by the National Natural Science Foundation of China under Grant 62371063, Grant 62471290, Grant 62321001, and Grant 92467301, in part by the Beijing Municipal Natural Science Foundation under Grant L232047, in part by Postdoctoral Fellowship Program of CPSF under Grant Number GZB20250810, in part by the China Postdoctoral Science Foundation under Grant Number 2025M783515, and in part by the Beijing Nova Program.}
		
		\thanks{Kailin Tan, Jincheng Dai, Kai Niu and Ping Zhang are with Beijing University of Posts and Telecommunications, Beijing 100876, China (e-mail: daijincheng@bupt.edu.cn).}
		
		\thanks{Sixian Wang, Guo Lu and Wenjun Zhang are with Shanghai Jiao Tong University, Shanghai 200240, China.}
		
		\thanks{Shuo Shao is with University of	Shanghai for Science and Technology, Shanghai 200093, China.}
		
		\thanks{Open-source code and data will be available on-line at: \url{https://github.com/semcomm/DiTJSCC}}
		
		\vspace{-1em}
	}
	
	\maketitle

	\begin{abstract}
		Generative joint source-channel coding (GenJSCC) has emerged as a new Deep JSCC paradigm for achieving high-fidelity and robust image transmission under extreme wireless channel conditions, such as ultra-low bandwidth and low signal-to-noise ratio. Recent studies commonly adopt diffusion models as generative decoders, but they frequently produce visually realistic results with limited semantic consistency. This limitation stems from a fundamental mismatch between reconstruction-oriented JSCC encoders and generative decoders, as the former lack explicit semantic discriminability and fail to provide reliable conditional cues. In this paper, we propose DiT-JSCC, a novel GenJSCC backbone that can jointly learn a semantics-prioritized representation encoder and a diffusion transformer (DiT)-based generative decoder. Our open-source project aims to promote future research in GenJSCC. Specifically, we design a semantics-detail dual-branch encoder that aligns naturally with a coarse-to-fine conditional DiT decoder, prioritizing semantic consistency under extreme channel conditions. Moreover, a training-free adaptive bandwidth allocation strategy inspired by Kolmogorov complexity is introduced to further improve transmission efficiency, thereby redefining the notion of information value in the era of generative decoding. Extensive experiments demonstrate that DiT-JSCC consistently outperforms existing JSCC methods in both semantic consistency and visual quality, particularly in extreme regimes.
	\end{abstract}
	

\IEEEpeerreviewmaketitle

\section{Introduction}\label{section_introduction}
\IEEEPARstart{R}ECENT advances in deep learning have revitalized the classic problem of joint source-channel coding (JSCC) \cite{fresia2010joint}, giving rise to a key enabler in semantic communication: Deep JSCC. For image transmission tasks, early Deep JSCC \cite{DJSCC, DJSCCF, DJSCCL} directly learns to map source image data to channel-input symbols using end-to-end learned autoencoders, optimizing toward pixel-wise distortion metrics, e.g., PSNR. Subsequent improvements \cite{yang2024swinjscc, dai2022nonlinear, wang2023improved} have demonstrated comparable or even superior rate-distortion performance compared to state-of-the-art engineered coding and transmission systems. 

\begin{figure}[t]
	\setlength{\abovecaptionskip}{0cm}
	\setlength{\belowcaptionskip}{0cm}
	\centering{\includegraphics[scale=0.29]{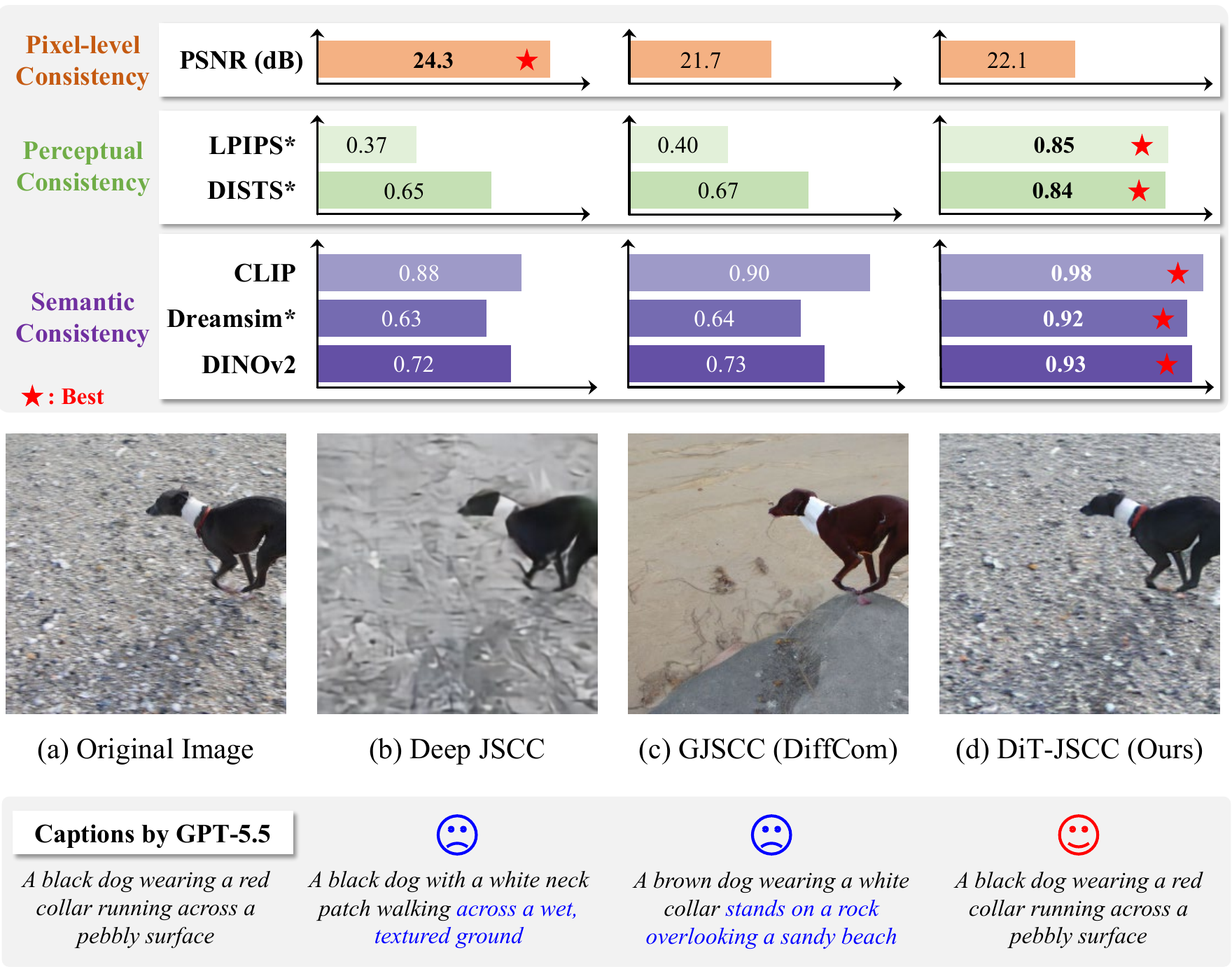}}
	\caption{Semantic consistency is more critical than pixel-level consistency for human perception under extremely limited bandwidth or low SNR conditions. SNR = 0 and CBR = 1/96 for all methods. Visually, our method produces the most faithful reconstructions, both realistic and semantically aligned with the original. In terms of metrics, our approach achieves the best performance across \add{perceptual consistency metrics (LPIPS \cite{lpips}, DISTS \cite{dists}) and semantic consistency metrics (Dreamsim \cite{NEURIPS2023_9f09f316}, CLIP \cite{radford2021learning}, DINOv2 \cite{oquab2023dinov2})}, while its PSNR is much lower than the original deep JSCC. Notably, only our method successfully captures the original textual semantics. * denotes a score inversion operation, for example, $\text{LPIPS*} = 1 - \text{LPIPS}$. \add{Thus, starred metrics such as LPIPS* and Dreamsim* follow the higher-is-better convention.}}
	\label{fig_sem_metrics}
	\vspace{0em}
\end{figure}

Despite these advancements, existing methods tend to produce noticeable blurry artifacts in reconstructed images that are misaligned with human visual quality (as exemplified in Fig. \ref{fig_sem_metrics} (b)). This issue becomes particularly noticeable under extreme wireless channel conditions, such as ultra-low bandwidth and low signal-to-noise ratio (SNR). The reason lies fundamentally in the lack of knowledge about the realistic data distribution in pixel-reconstruction-based decoders. \add{In this context, as an upgrade of Deep JSCC,} Generative JSCC (GenJSCC) \cite{yang2022ofdm, erdemir2023generative, tan2024rate} has now emerged as a promising paradigm toward robust and high-fidelity end-to-end transmission under challenging channel conditions. The core idea of this paradigm leverages generative priors and stochastic sampling to perform generative decoding, effectively mitigating the random perturbations introduced by wireless channels.

A recent trend in GenJSCC is the adoption of diffusion models \cite{ho2020denoising} as powerful generative decoders at the receiver \cite{niu2023hybrid, wu2024cddm, zhang2025semantics, yang2024diffusion, wang2025diffcom}. By leveraging rich generative priors, these approaches can synthesize visually realistic reconstructions even under extreme channel conditions, thereby substantially enhancing perceptual quality. However, existing diffusion-based GenJSCC methods exhibit a prevalent limitation in semantic consistency: the reconstructed images often contain discrepancies in local structures or even object categories compared with the original inputs. As illustrated in Fig. \ref{fig_sem_metrics}, from an objective evaluation perspective, conventional pixel-wise metrics such as PSNR become unreliable for assessing semantic consistency under such conditions, due to the inherent diversity of outputs produced by diffusion decoders. In contrast, \add{feature-based perceptual consistency metrics, including LPIPS \cite{lpips} and DISTS \cite{dists}, together with semantic consistency metrics such as DreamSim, CLIP, and DINOv2 \cite{NEURIPS2023_9f09f316,radford2021learning,oquab2023dinov2},} are more appropriate, as they operate in latent feature spaces and are robust to pixel-level misalignments. Consequently, under such extreme channel conditions, semantic consistency is increasingly regarded as the primary criterion for evaluating end-to-end image transmission performance, while appearance-level detail plays a secondary role.

Motivated by the increasing importance of semantic consistency under extreme channel conditions, we challenge these conventional GenJSCC designs by systematically rethinking what information the encoder should extract and transmit, and how much of it is required, to reliably support the conditional generative decoding \cite{zhang2023adding, chen2024pixart}. Despite the empirical success of existing GenJSCC approaches \cite{niu2023hybrid, wu2024cddm, zhang2025semantics, yang2024diffusion, wang2025diffcom}, there remains limited understanding of the underlying mechanisms by which transmitted representations effectively and robustly guide conditional sampling in diffusion-based decoders. A fundamental question therefore arises: is diffusion-based decoding primarily driven by reusing legacy, pixel-reconstruction-oriented encoders, as commonly adopted in existing GenJSCC methods, or does it instead rely on a carefully designed encoder that explicitly decouples semantic and detail information, with a greater portion of bandwidth allocated to semantically salient representations?

Understanding these mechanisms is critical for advancing Generative JSCC, as they directly determine how to select appropriate target representations and maximize their effectiveness in guiding stochastic sampling within generative decoders. A prevailing insight from the image generation community is that effective diffusion guidance is strongly correlated with global semantic understanding. In particular, high-level semantic representations encoded by large vision foundation models (VFMs), such as MAE \cite{he2022masked}, JEPA \cite{assran2023self}, and DINOv2 \cite{oquab2023dinov2}, have been shown to provide more efficient and robust guidance for generating samples with strong semantic consistency. Consequently, emphasizing the transmission of high-level semantic features as the primary information carrier is essential for the design of effective GenJSCC frameworks.

Our key insight, supported by empirical evidence \cite{singh2025matters}, is that high-level semantic representations extracted from VFM encoders primarily capture the basic structures of images, which play a crucial role in reconstruction fidelity, especially under extreme bandwidth constraints and low SNR regimes. In contrast, low-level details, typically comprising high-frequency information such as textures and edges, are less important from a generative perspective. 
Building on the above insight, in this paper, we propose a novel semantics-prioritized GenJSCC framework, named DiT-JSCC. At the transmitter, we develop a dual-branch JSCC encoder, which disentangles the input images hierarchically into high-level semantics and low-level details. At the receiver, we employ a generative diffusion transformer (DiT) model as the backbone of the JSCC decoder to synthesize missing details and render them based on recovered semantics. In this manner, our DiT-JSCC co-designs the transceiver and enables end-to-end optimization, which differs from existing diffusion-based GenJSCC methods \cite{niu2023hybrid, wu2024cddm, zhang2025semantics, yang2024diffusion, wang2025diffcom} that primarily emphasize the denoising generative process at the receiver.

To realize the dual-branch DiT-JSCC, we need to address two challenges: (1) how to allocate bandwidth between the semantic and detail branches effectively, and (2) how to leverage multi-level guidance efficiently throughout the diffusion denoising process. For the first challenge, inspired by recent VFM-driven diffusion works \cite{yu2025representation, leng2025repa, yao2025reconstruction}, we adopt a pretrained VFM, DINOv2 \cite{oquab2023dinov2}, to extract high-level semantic features. After that, we develop a semantics-prioritized bandwidth allocation (BA) strategy, i.e., first satisfy the bandwidth demand of the semantic branch and then allocate the remaining budget to the detail branch. To quantitatively estimate the required semantic bandwidth in a content-adaptive manner, we draw inspiration from Kolmogorov complexity (KC) \cite{li2008introduction}, positing that the semantic cost should scale with the descriptive complexity of the input image. Since KC is an idealized and fundamentally non-computable quantity \cite{li2008introduction}, we employ image captions as a tractable surrogate and further develop a hybrid textual complexity estimation algorithm. For the second challenge, we propose to fuse multi-level guidance at different depths of the diffusion denoising network, yielding a coarse-to-fine generation paradigm: high-level semantic guidance dominates early layers to stabilize global structure, while low-level detail guidance is injected progressively in later layers to refine textures and edges. We implement the JSCC decoder with a DiT backbone \cite{bao2023all, ma2024sit, peebles2023scalable}, which is the de facto mainstream architecture for high-quality diffusion generation.

The remainder of this paper is organized as follows. In Section \ref{section_system_model}, we compare DiT-JSCC with conventional Deep JSCC and GenJSCC approaches in terms of system architecture. Section \ref{section_method} presents the implementation details of DiT-JSCC. Experimental results are presented in Section \ref{section_experimental_results}, where we quantitatively compare our method with several baselines to demonstrate its performance advantages. Finally, Section \ref{section_conclusion} concludes the paper.

\begin{figure}[t]
	\setlength{\abovecaptionskip}{0cm}
	\setlength{\belowcaptionskip}{0cm}
	\centering{\includegraphics[scale=0.54]{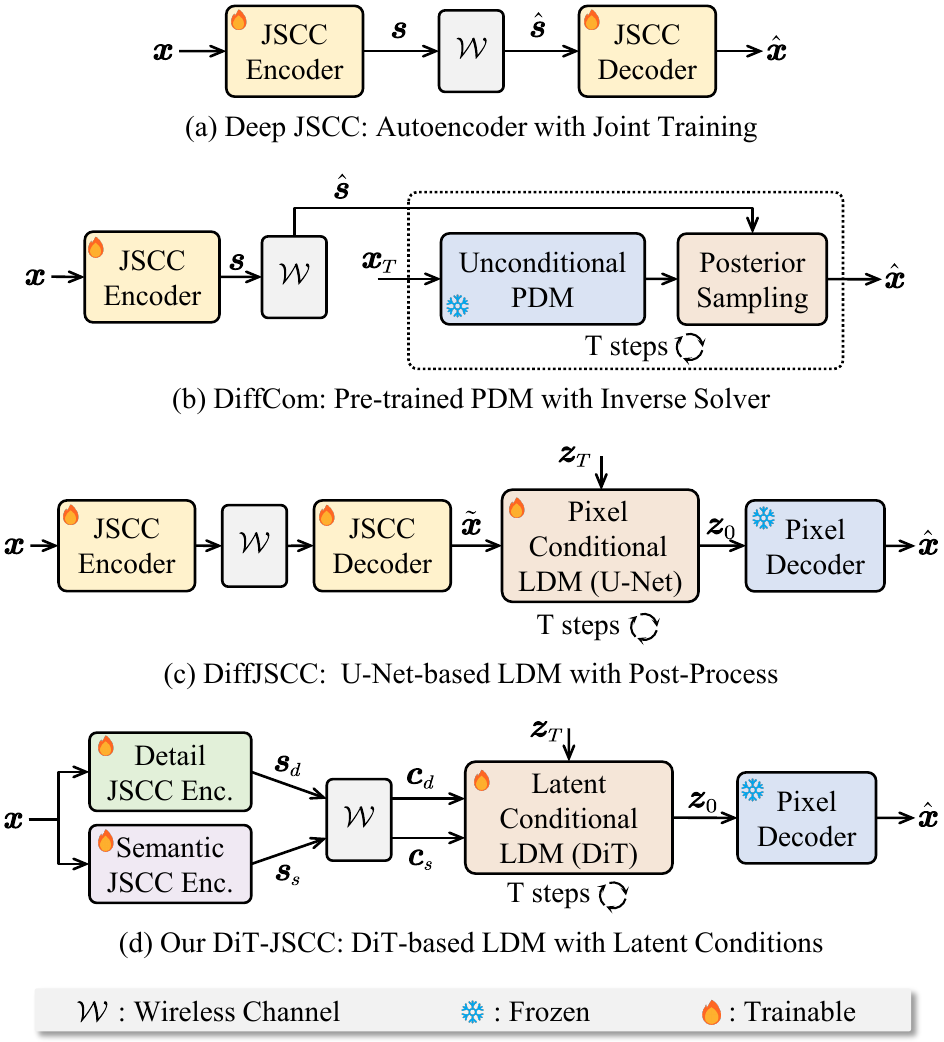}}
	\caption{Overall framework of the proposed DiT-JSCC (d) and comparison with the original deep JSCC (a) and other existing Generative JSCC structures (b and c).}
	\label{fig_framework_comparison}
	\vspace{0em}
\end{figure}

%\begin{figure}[t]
%	\setlength{\abovecaptionskip}{0cm}
%	\setlength{\belowcaptionskip}{0cm}
%	\centering{\includegraphics[scale=0.29]{figure_PPT/fig2_1.pdf}}
%	\caption{Illustration of information degradation behavior under different encoders as the bandwidth decreases. (a) The Reconstruction-based encoder learns a mixed representation, showing no explicit preference between high-level semantics and low-level details. (b) Our method explicitly separates them and controls the priority retention of high-level semantics, which are more valuable for the generative decoder. The quantitative analysis can be found in Fig. \ref{fig13}.}
%	\label{fig2_1}
%	\vspace{0em}
%\end{figure}

\section{System Model} \label{section_system_model}

\subsection{Preliminaries of Deep JSCC}
We consider the problem of transmitting a RGB image $\boldsymbol{x}\in \mathbb{R} ^{H\times W\times 3}$ over a point-to-point wireless channel, where $H$ and $W$ denote the height and width of the image, respectively. Under the original Deep JSCC setup \cite{DJSCC, DJSCCF, DJSCCL}, a JSCC encoder $\mathcal{E}$ encodes the image $\bm{x}$ into a vector of complex-valued channel input symbols $\boldsymbol{s}\in \mathbb{C}^k$. After power normalization which ensures an average power constraint, these $k$ symbols are transmitted over a noisy wireless channel, and the receiver gets a sequence $\boldsymbol{\hat{s}}=\mathcal{W}\left( \boldsymbol{s} \right) $. In this paper, we consider the general fading channel model such that the transfer function is $\boldsymbol{\hat{s}}=\mathcal{W}\left( \boldsymbol{s} \right) =\boldsymbol{h}\odot \boldsymbol{s}+\boldsymbol{n}$ where $\odot$ is the element-wise product, $\bm{h}$ is the channel gain vector, and each component of the noise $\bm{n}$ is independently sampled from a Gaussian distributions, i.e., $\boldsymbol{n}\sim p_{\boldsymbol{n}}\triangleq \mathcal{N} \left( 0,\sigma _{\boldsymbol{n}}^{2}\boldsymbol{I} \right)$, where $\sigma _{\boldsymbol{n}}^{2}$ is noise power. For AWGN channel, $\bm{h}$ is set to 1. At the receiver, a symmetrical decoder $\mathcal{D}$ is employed to reconstruct the image from the received signals $\boldsymbol{\hat{s}}$. The system efficiency is quantified by the channel bandwidth ratio (CBR), which is referred as $\rho$ and is defined as:
\begin{equation}\label{CBR}
	\begin{aligned}
		\rho =\frac{k}{3\times H\times W}.
	\end{aligned}
\end{equation}
It represents the average number of available channel symbols per source dimension.

The goal of this system is to maximize the consistency between the input and reconstructed images under a given bandwidth constraint. Typically, $\mathcal{E}$, $\mathcal{D}$, and the parameter-free wireless channel are jointly optimized to minimize the end-to-end distortion objective under a constant bandwidth rate constraint $\rho_c$, formulated as:
\begin{equation}\label{Distortion_loss}
	\begin{aligned}
		\mathbb{E} _{\boldsymbol{x}\sim p\left( \boldsymbol{x} \right) ,\boldsymbol{h}\sim p\left( \boldsymbol{h} \right) ,\boldsymbol{n}\sim p\left( \boldsymbol{n} \right)}\left[ \mathcal{L} _D\left( \boldsymbol{x}, \boldsymbol{\hat{x}} \right) \right], \text{with constant~} \rho_c, 
	\end{aligned}
\end{equation}
where $\mathcal{L} _D$ denotes a distance measure typically formulated by pixel-level distortions such
as mean-squared error (MSE), and $\rho_c$ is determined by the autoencoder bottleneck dimension.

However, a common limitation of existing Deep JSCC approaches is that their reconstructions often suffer from clear blurry artifacts, which are misaligned with human visual perception. This issue becomes particularly pronounced under extreme channel conditions, such as ultra-low bandwidth and low SNR. The root cause lies in the reconstruction-oriented training objective, which biases the model toward learning the pixel-wise average of plausible solutions across the dataset. Consequently, the decoder lacks prior knowledge about the realistic data distribution. In contrast, perceptual quality is more accurately characterized by the divergence between the distributions of the source and reconstructed data \cite{blau2019rethinking}. 

\begin{figure*}[t]
	\setlength{\abovecaptionskip}{0cm}
	\setlength{\belowcaptionskip}{0cm}
	\centering{\includegraphics[scale=0.58]{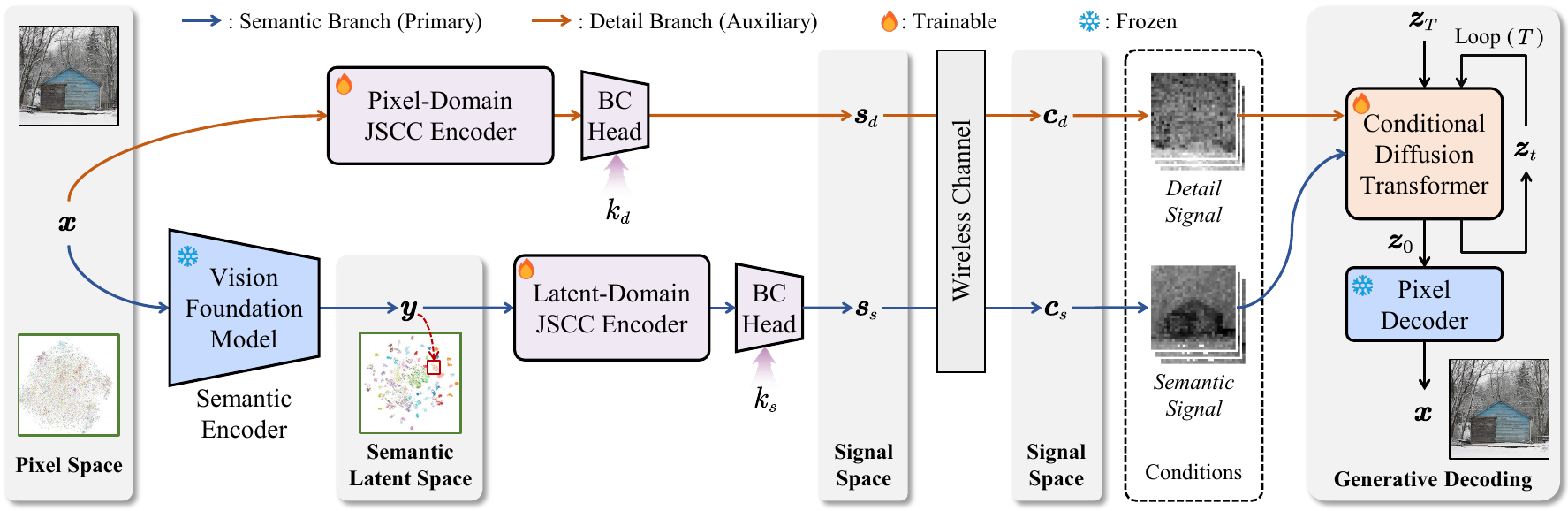}}
	\caption{ Overview of our DiT-JSCC system architecture.}
	\label{fig_system_architecture}
	\vspace{0em}
\end{figure*}
\subsection{Diffusion-based Generative JSCC}
In recent years, diffusion models have been increasingly adopted to enhance the perceptual quality in Deep JSCC systems, giving rise to a new class of Generative JSCC methods \cite{niu2023hybrid, wu2024cddm, zhang2025semantics, yang2024diffusion, wang2025diffcom}. In Fig. \ref{fig_framework_comparison}, we illustrate two advanced methods following a conditional generative decoding paradigm. DiffCom \cite{wang2025diffcom} directly utilizes the raw channel-received signal $\bm{\hat{s}}$ as a fine-grained condition to guide score-matching posterior sampling using a frozen unconditional pixel-domain diffusion model (PDM). DiffJSCC \cite{yang2024diffusion} extracts textual and visual features from the lossy reconstructions $\boldsymbol{\tilde{x}}$ produced by the JSCC autoencoder, and trains a U-Net-based conditional latent diffusion model (LDM) to generate improved reconstructions $\boldsymbol{\tilde{x}}$. Another paradigm treats the received signal as an intermediate state along a diffusion trajectory and learns a dedicated channel denoiser to restore the original channel input. However, works following the latter paradigm are currently limited to low-resolution images (e.g., $128 \times 128$), and are thus not included in our comparative analysis. In summary, all these Generative JSCC methods leverage powerful priors for generative decoding to compensate for the realism limitations of original Deep JSCC, achieving significantly improved FID \cite{fid} scores.

However, we observe a severe decline in semantic consistency under extreme channel conditions for these methods: their reconstructed images $\bm{\hat{x}}$ often deviate from the original input $\bm{x}$ in terms of local structures or even object categories. This issue is also reflected in various \add{objective perceptual and semantic consistency metrics}. We attribute this issue to two primary factors:
\begin{enumerate}
	\item \emph{Insufficient semantic preservation:} Current GenJSCC methods reuse the JSCC encoder learned by \eqref{Distortion_loss} under the reconstruction tasks, where the transmitted signal $\bm{z}$ does not sufficiently retain high-level semantics. Especially, DiffJSCC adopt a post-process approach to refine the distorted results, introducing further information loss.
	
	\item \emph{Inadequate conditional guidance:} DiffCom with the inverse solver only guides the posterior sampling process, lacking strong constraints on the unconditional prior. DiffJSCC merely fine-tunes the conditional LDM without joint optimization with the encoder side, leading to limited effectiveness in capturing semantics-relevant conditions.
\end{enumerate}
We argue that addressing this issue requires two key improvements: (i) redesigning a semantics-oriented encoder to generate high-quality semantic representations for generative decoding, and (ii) establishing tight collaboration between the encoder and decoder in both architectural design and joint training, thereby enabling stronger conditional guidance throughout the generation process.

\subsection{Semantics-Prioritized Generative JSCC}
In this paper, we propose a novel Generative JSCC framework, called DiT-JSCC, which prioritizes the transmission of high-level semantics for conditional generative decoding under extreme channel conditions. As illustrated in Fig. \ref{fig_framework_comparison}, we introduce a dual-branch JSCC encoding structure to explicitly decompose high-level semantics and low-level details. For reconstruction-based JSCC encoders, these two components are inherently entangled within the encoded representations and tend to degrade in a similar fashion as bandwidth decreases, without a clear semantic-first transmission order. Specifically, in our framework, the primary semantic JSCC encoder compresses and transmits the semantic signal $\bm{s}_s$, while the auxiliary detail JSCC encoder captures residual details and encodes them into a complementary detail signal $\bm{s}_d$.

At the receiver, the recovered latent signals $\bm{c}_s$ and $\bm{c}_d$ are directly used to condition a latent diffusion model (LDM) built upon a powerful DiT backbone. Within this framework, the encoder, wireless channel, and conditional DiT decoder are optimized jointly, enabling tight coupling between the communication and generative processes. Such joint optimization improves the semantic alignment of the conditioning representations and allows the diffusion model to better accommodate channel-induced perturbations, thereby enhancing the robustness of generative decoding. The following section details the implementation of the proposed DiT-JSCC framework.

\section{Methodology}\label{section_method}

The whole system architecture of the proposed DiT-JSCC is depicted in Fig. \ref{fig_system_architecture}. It contains two key components: the VFM-driven dual-branch JSCC encoding structure and the coarse-to-fine conditional DiT (CDiT) decoder. In addition, we introduce a KC-inspired bandwidth allocation strategy to further enhance transmission efficiency in section \ref{sectionIII_C}.

\subsection{\texorpdfstring{VFM-Driven Dual-Branch JSCC Encoding Structure}{VFM-Driven Dual-Branch JSCC Encoding Structure}} \label{sectionIII_A}

%To enable semantics-prioritized transmission, our dual-branch JSCC encoding structure must not only ensure functional complementarity between the two branches, but also allow for flexible bandwidth control for each branch. To achieve this, inspired by VFM-driven diffusion works \cite{yu2025representation, leng2025repa, yao2025reconstruction}, we incorporate a pretrained VFM (DINOv2 \cite{oquab2023dinov2}) into our encoder and keep it frozen during training. This allows our framework to directly inherit the powerful semantic representation capabilities of VFM, providing a solid foundation for subsequent architecture design and optimization.

To enable semantics-prioritized transmission, the proposed dual-branch JSCC encoder is designed to ensure functional complementarity between the two branches while allowing flexible and independent bandwidth control. Inspired by recent VFM-driven diffusion models \cite{yu2025representation, leng2025repa, yao2025reconstruction}, we incorporate a pretrained VFM, specifically DINOv2 \cite{oquab2023dinov2}, into the encoder and keep it frozen during training. By doing so, the proposed framework directly inherits the strong semantic representation capability of the VFM, establishing a robust and semantically meaningful foundation for subsequent architectural design and optimization.

\subsubsection{\texorpdfstring{VFM-Driven Semantic Branch}{VFM-Driven Semantic Branch}} 
In the semantic branch, we first employ a pre-trained VFM $E_{\mathrm{VFM}}$ to map the image $\boldsymbol{x}\in \mathbb{R} ^{H\times W\times 3}$ from the pixel space into a compact, semantically rich latent space, resulting in the semantic representation $\boldsymbol{y}\in \mathbb{R} ^{\frac{H}{16}\times \frac{W}{16}\times 256}$. The VFM trained via self-supervised learning, constructs a latent space with clear semantic separation and strong discriminative structure. We emphasize that pixel-domain JSCC encoders trained from scratch under transmission tasks struggle to achieve this.

Then we cascade a latent-domain JSCC (LD-JSCC) encoder $f_\mathrm{LD}$ with a Bandwidth-Control (BC) Head $\mathcal{H}_s$ to maps $\bm{y}$ into channel-input signal $\bm{s}_s$ consisting of $k_s$ channel symbols. The LD-JSCC encoder is adapted from the SwinJSCC architecture \cite{yang2024swinjscc}, with modifications to perform compression only along the channel dimension in order to maintain spatial alignment. The BC Head, inspired by the Rate ModNet \cite{yang2024swinjscc}, enables flexible bandwidth control by adjusting $k_s$. On the receiver side, the received semantic signal $\bm{c}_s$ are directly used to guide the generative decoding process. The overall process in the semantic branch can be formulated as:
\begin{equation}\label{Branch_semantic}
	\begin{aligned}
		\boldsymbol{x}\xrightarrow{E_{\mathrm{VFM}}(\cdot )}\boldsymbol{y}\xrightarrow{\mathcal{H} _s\left( f_{\mathrm{LD}}(\cdot ),k_s \right)}\boldsymbol{s}_s\xrightarrow{\mathcal{W} (\cdot ,\boldsymbol{h})}\boldsymbol{c}_s.
	\end{aligned}
\end{equation}

\begin{figure}[t]
	\setlength{\abovecaptionskip}{0cm}
	\setlength{\belowcaptionskip}{0cm}
	\centering{\includegraphics[scale=0.32]{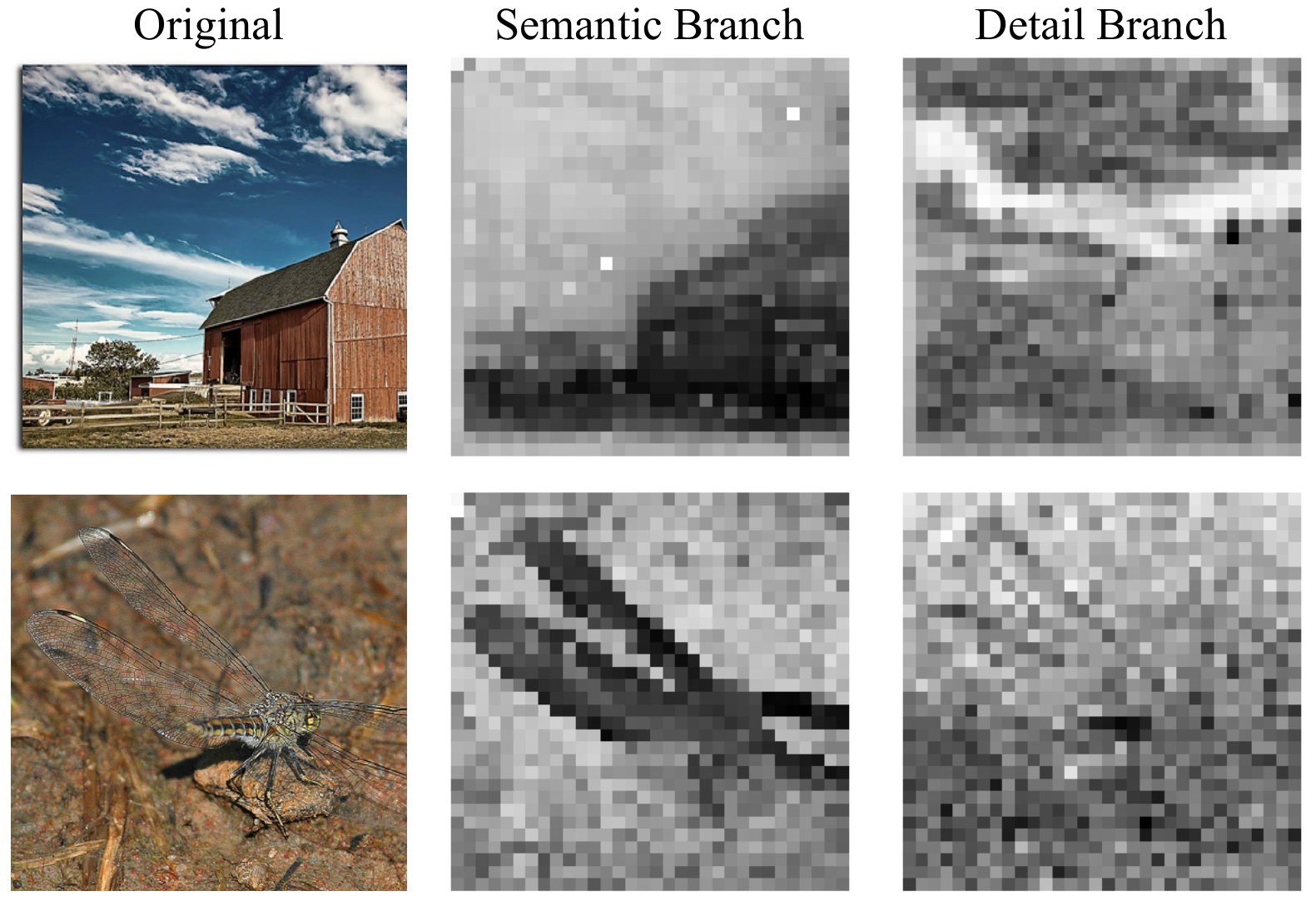}}
	\caption{The representations learned from different branches exhibit distinct characteristics: the semantic branch focuses on structurally regular and semantically rich objects, while the detail branch targets high-frequency regions with details.}
	\label{fig_representation_maps}
	\vspace{0em}
\end{figure}

\subsubsection{Detail Branch}
Given the limitations of VFMs in capturing fine-grained details, we introduce a detail branch as a auxiliary branch. Specifically, we employ a pixel-domain JSCC (PD-JSCC) encoder $f_\mathrm{PD}$ followed by a BC Head  $\mathcal{H}_d$ to directly map the pixel-level $\bm{x}$ into the signal space, yielding the channel-input detail signal $\bm{s}_d$ of $k_d$ channel symbols. The PD-JSCC encoder shares the same network architecture as the SwinJSCC encoder: it first performs four stages of spatial downsampling and then compresses along the channel dimension. After traversing the wireless channel, the received detail signal $\bm{c}_d$ with $\bm{c}_s$ jointly guide the generative decoding process. The detail branch can be formulated as:
\begin{equation}\label{Branch_Detail}
	\begin{aligned}
		\boldsymbol{x}\xrightarrow{\mathcal{H}_d \left( f_{\mathrm{PD}}(\cdot ),k_d \right)}\boldsymbol{s}_d\xrightarrow{\mathcal{W} (\cdot ,\boldsymbol{h})}\boldsymbol{c}_d.
	\end{aligned}
\end{equation}
By joint training, the detail branch is forced to learn the residual detail features that are not captured by the semantic branch.

\subsubsection{Analysis on Representations and Bandwidth of Dual Branches}
We first analyze the representational characteristics of the two branches. As illustrated in Fig. \ref{fig_representation_maps}, we visualize the intermediate representation maps from both the semantic and detail branches. It is evident that the semantic branch primarily captures the main semantic objects in the image (those with regular structures or clear semantic meaning), while relatively less emphasis on semantically simpler or ambiguous content. Differently, the detail branch focuses on regions rich in details, corresponding to high-frequency components, and it lacks clear semantic discriminability. These observations suggest that the two branches offer complementary functions in the transmission of image information.

\begin{figure}[t]
	\setlength{\abovecaptionskip}{0cm}
	\setlength{\belowcaptionskip}{0cm}
	\centering{\includegraphics[scale=0.26]{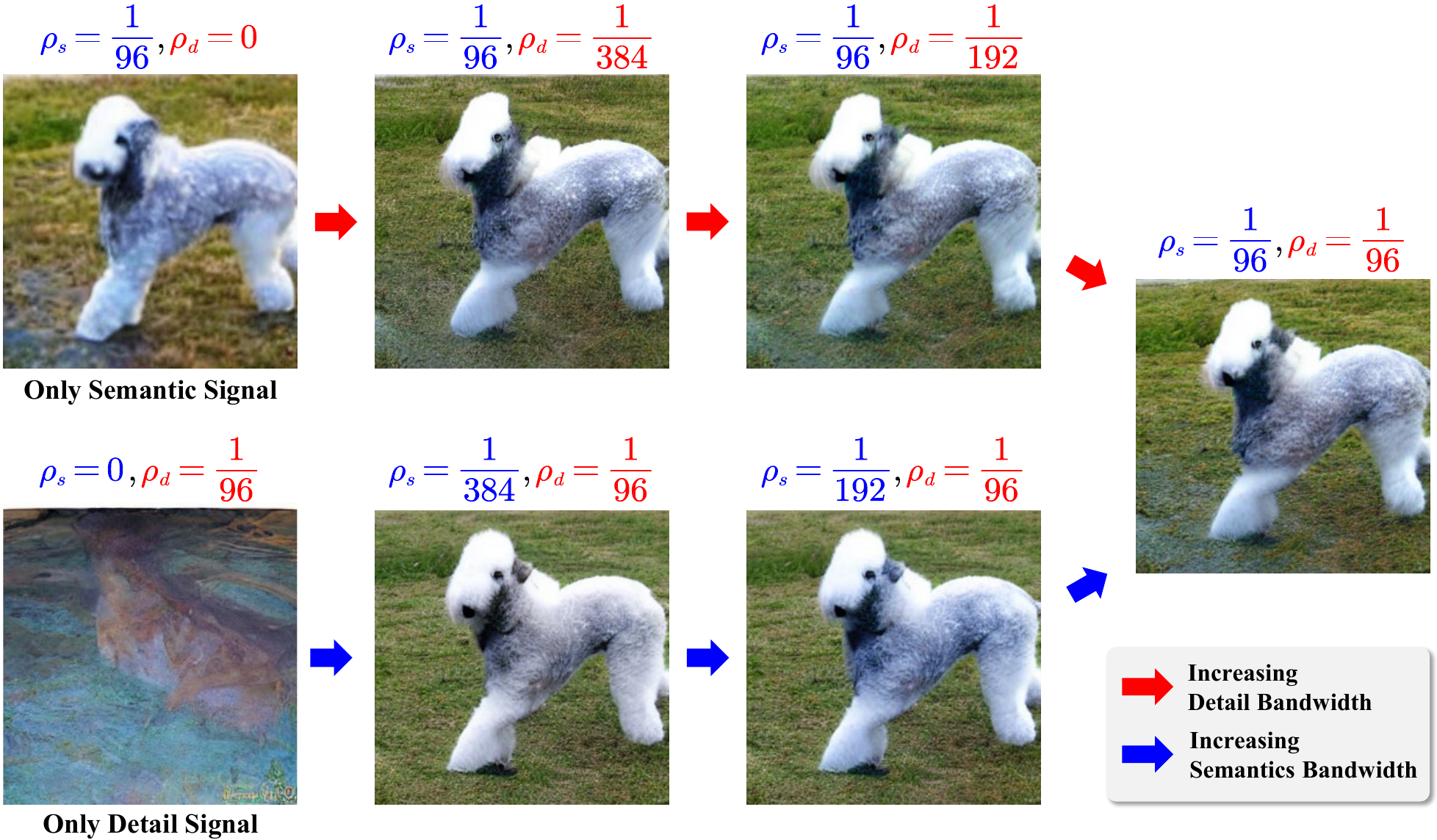}}
	\caption{Reconstruction results under different bandwidth allocations of the two branches. When only the semantic signal is available ($\rho _s>0, \rho _d=0$), the reconstruction retains the overall semantic structure but appears blurry details. In contrast, when only the detail signal is transmitted ($\rho _s=0, \rho _d>0$), the model fails to produce meaningful reconstructions.}
	\label{fig_visual_results4dual_branches}
	\vspace{0em}
\end{figure}

Next, we examine how the bandwidth of two branches affects reconstruction performance. In our framework, the bandwidth allocated to each branch can be flexibly controlled through two tunable hyperparameters, $k_s$ and $k_d$, respectively. The CBR for each branch can be computed as:
\begin{equation}\label{E_Detail}
	\begin{aligned}
		\rho _s=\frac{k_s}{3\times H\times W}, \,\,\,\, \rho _d=\frac{k_d}{3\times H\times W}.
	\end{aligned}
\end{equation}
As shown in Fig. \ref{fig_visual_results4dual_branches}, when only semantic signal is provided ($\rho _s>0, \rho _d=0$), DiT-JSCC successfully reconstructs the semantic content but with blurred details. As the bandwidth of the detail branch increases, fine-grained details are progressively restored. However, in the reverse setting where only the detail signal is transmitted ($\rho _s=0, \rho _d>0$), the model fails to reconstruct a meaningful image. These results verify the complementary roles of the two branches and reinforce the design principle of our framework: semantic branch should be prioritized to ensure meaningful reconstruction.

\begin{figure*}[t]
	\setlength{\abovecaptionskip}{0cm}
	\setlength{\belowcaptionskip}{0cm}
	\centering{\includegraphics[scale=0.6]{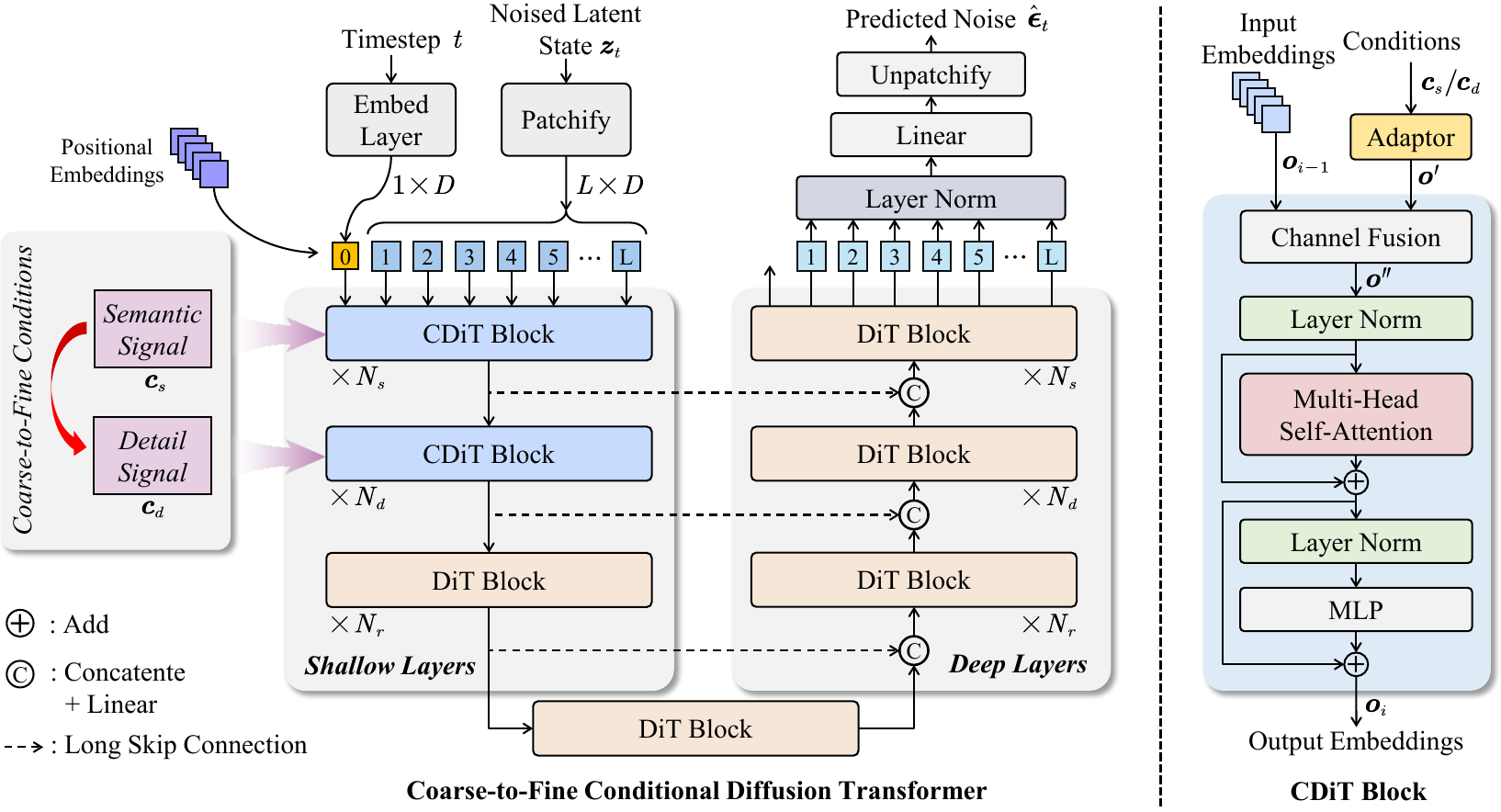}}
	\caption{The overview of the proposed coarse-to-fine CDiT model. The left part illustrates the network architecture incorporates a coarse-to-fine conditional injection strategy and a U-Net-like backbone. The right part presents the network details of the CDiT block.}
	\label{fig_CDiT_network}
	\vspace{0em}
\end{figure*}

\subsection{ Coarse-to-Fine Conditional Diffusion Transformer}  \label{sectionIII_B}
The generative decoding paradigm leverages stochastic sampling to provide rich prior knowledge and generate diverse outputs. In our setting of using conditional guidance, accurately guiding the diffusion decoding process is critical to achieving high-level semantic consistency under severely degraded channel conditions. In this subsection, we focus on designing a conditional LDM that matches with the proposed dual-branch encoding structure, thereby facilitating effective transceiver collaboration.

To this end, we employ emerging DiT \cite{bao2023all, ma2024sit, peebles2023scalable} as the backbone architecture and design a coarse-to-fine conditional DiT (CDiT) model for generative decoding. As illustrated in Fig. \ref{fig_system_architecture}, our CDiT decoder is conditioned on the received semantic signal $\bm{c}_s$ and detail signal $\bm{c}_d$. The goal is to provide coarse-to-fine control over the generation process, allowing the decoder to progressively refine the noisy latent $\bm{z}_t$ in alignment with both semantic intent and local textures. Compared with generative decoding methods that rely on texts or sketches, our method provides impact and integrated semantic cues. Furthermore, by jointly training with the wireless channel, our conditional signals are inherently more robust to channel perturbations. In contrast, methods based on texts or sketches typically require lossless transmission of the two explicit conditional signals; otherwise, their performance degrades significantly.

In the remainder of this subsection, we first provide a brief overview of the LDM as a preliminary, followed by implementation details of the proposed CDiT architecture.

\subsubsection{Preliminary of LDM}
To enhance training stability and reduce the computational burden, we adopt the LDM framework \cite{rombach2022high}. Instead of performing diffusion directly in the pixel space, LDM applies the Denoising Diffusion Probabilistic Model (DDPM) \cite{ho2020denoising} in a learned latent space. Specifically, a variational autoencoder (VAE) first projects the input image $\bm{x}$ into a compact latent representation $\bm{z}_0$, which is then progressively noised and denoised in the latent space. During the forward process, Gaussian noise is incrementally added to the clean latent representation $\bm{z}_0$, with the noise strength modulated by a predefined noise schedule $\beta _t$. This forward diffusion process can be formulated as:
\begin{equation}\label{SC}
	\begin{aligned}
		\boldsymbol{z}_t=\sqrt{\bar{\alpha}_t}\boldsymbol{z}_0+\sqrt{1-\bar{\alpha}_t}\boldsymbol{\epsilon }
	\end{aligned}
\end{equation}
where $\boldsymbol{\epsilon }\sim \mathcal{N} \left( 0,\mathbf{I} \right)$ is a sample from a standard Gaussian distribution, and $t$ is the time step. Here, $\alpha _t=1-\beta _t$ and $\bar{\alpha}_t=\prod\nolimits_{i=1}^t{\alpha _i}$.

The denoising process serves to progressively reconstruct the clean latent representation $\bm{z}_0$ from the fully perturbed latent variable $\bm{z}_T$ through a sequence of $T$ reverse steps. As established in \cite{ho2020denoising}, the conditional probability $p_{\theta}(\bm{z}_{t-1}|\bm{z}_t)$ in the reverse Markov chain can be learned by minimizing the variational Evidence Lower Bound (ELBO), which is mathematically equivalent to minimizing the Kullback-Leibler (KL) divergence between the learned reverse distribution $p_{\theta}(\bm{z}_{t-1}|\bm{z}_t)$ and the exact posterior $q(\bm{z}_{t-1}|\bm{z}_t,\bm{z}_0)$. This posterior distribution $q(\bm{z}_{t-1}|\bm{z}_t,\bm{z}_0)$ is tractable and can be analytically expressed as a Gaussian distribution:
\begin{equation}\label{SC}
	\begin{aligned}
		q(\boldsymbol{z}_{t-1}|\boldsymbol{z}_t,\boldsymbol{z}_0)=\mathcal{N} \left( \boldsymbol{z}_{t-1};\mu _t\left( \boldsymbol{z}_t,\boldsymbol{z}_0 \right) ,\sigma _{t}^{2}\mathbf{I} \right) ,
	\end{aligned}
\end{equation} 
where the mean $\mu_t(\bm{z}_t, \bm{z}_0)$ is defined as:
\begin{equation}\label{SC}
	\begin{aligned}
		\mu _t\left( \boldsymbol{z}_t,\boldsymbol{z}_0 \right) =\frac{1}{\sqrt{\alpha _t}}\left( \boldsymbol{z}_t-\frac{1-\alpha _t}{\sqrt{1-\bar{\alpha}_t}}\boldsymbol{\epsilon } \right) .
	\end{aligned}
\end{equation} 
To approximate this posterior during generation, the model defines $p_\theta(\bm{z}_{t-1}|\bm{z}_t)$ as a Gaussian:
\begin{equation}\label{SC}
	\begin{aligned}
		p_{\theta}(\boldsymbol{z}_{t-1}|\boldsymbol{z}_t)=\mathcal{N} \left( \boldsymbol{z}_{t-1};\mu _{\theta}\left( \boldsymbol{z}_t,t \right) ,\sigma _{t}^{2}\mathbf{I} \right),
	\end{aligned}
\end{equation} 
with the mean parameterized by a neural network as:
\begin{equation}\label{SC}
	\begin{aligned}
		\mu _{\theta}\left( \boldsymbol{z}_t,t \right) =\frac{1}{\sqrt{\alpha _t}}\left( \boldsymbol{z}_t-\frac{1-\alpha _t}{\sqrt{1-\bar{\alpha}_t}}\boldsymbol{\epsilon }_{\theta}\left( \boldsymbol{z}_t,t \right) \right) ,
	\end{aligned}
\end{equation} 

Here, the core training objective reduces to accurately estimating the noise term $\bm{\epsilon}$ added during the forward diffusion. This estimation is accomplished via a neural network $\bm{\epsilon}_\theta$, which can be conditioned on auxiliary context variables $\bm{c}$ (e.g., text embeddings, latent features, etc.). Accordingly, the training loss is formulated as:
\begin{equation}\label{loss_DM}
	\begin{aligned}
		\mathcal{L} =\mathbb{E} _{\boldsymbol{z}_0,\boldsymbol{c},t,\boldsymbol{\epsilon }}\left\| \boldsymbol{\epsilon }-\boldsymbol{\epsilon }_{\theta}\left( \sqrt{\bar{\alpha}_t}\boldsymbol{z}_0+\sqrt{1-\bar{\alpha}_t}\boldsymbol{\epsilon },\boldsymbol{c},t \right) \right\| _{2}^{2}.
	\end{aligned}
\end{equation} 
This objective enables the denoising network to learn an effective approximation of the corruption process in reverse. Over successive iterations, the model becomes capable of reliably recovering clean representations from severely degraded inputs, guiding the generative process toward faithful and high-fidelity reconstructions.

\subsubsection{The Architecture of Coarse-to-Fine CDiT}
The architecture of the proposed coarse-to-fine CDiT (i.e., the denoising network $\bm{\epsilon}_\theta$) is shown in Fig. \ref{fig_CDiT_network}. It adopts a DiT backbone augmented with a U-Net-like architecture to accelerate training convergence and improve stability \cite{bao2023all}. More importantly, on this basis, we introduce a novel coarse-to-fine conditioning strategy to enable more effective integration of the DiT backbone with the two types of signals.

At each denoising step, CDiT takes as input the noised latent state $\boldsymbol{z}_t$, the timestep $t$, and a set of coarse-to-fine conditions to predict the corresponding noise component. Specifically, the spatial input $\boldsymbol{z}_t$ is first converted into a token sequence of length $L$ and dimension $D$ via a patchify layer, where $L$ corresponds to the number of spatial patches. The timestep $t$ is embedded into a single token using a sinusoidal positional encoding followed by an MLP projection. These two inputs are then concatenated to form an extended token sequence $\bm{o}_0$ of length $L+1$. To learn spatial positions, $\bm{o}_0$ is further augmented with learnable frequency-based positional embeddings before being fed into the DiT backbone. The DiT backbone comprises $2N+1$ CDiT/DiT blocks, organized into three stages: the ($N+1$)-th block serves as the middle layer, and the remaining $N$ blocks serve as the \emph{deep layers}. To facilitate gradient propagation and preserve hierarchical information, we integrate long skip connections between symmetric blocks across the shallow and deep layers, resulting in a symmetric U-Net-like structure.

In the coarse-to-fine conditioning strategy, semantic and detail signals are injected into distinct CDiT blocks independently. Inspired by \cite{yu2025representation}, which reveals that earlier DiT blocks are more responsive to high-level semantic representations while later blocks focus on low-level visual details, we align the granularity of conditional information with the hierarchical representation learning behavior of DiT. Specifically, $\bm{c}_s$ is injected into the first $N_s$ CDiT blocks, while $\bm{c}_d$ into the subsequent blocks. \add{In our implementation, the conditional CDiT blocks are placed in the shallow layers so that the received semantic and detail cues guide the denoising process before symmetric deep-layer refinement.} %This targeted injection improves conditional control. %by synchronizing the information content with the progressive representation learning of the denoising process.

Consequently, the shallow layers comprise $N_c=N_s+N_d$ CDiT blocks and $N_r$ standard DiT blocks. The complete backbone can thus be formalized as:
\begin{equation}\label{SC}
	\begin{aligned}
		\boldsymbol{o}_i=\begin{cases}
			\mathrm{B}_i\left( \boldsymbol{o}_{i-1}, \boldsymbol{c}_s \right)&		i=1, \cdots ,N_s\\
			\mathrm{B}_i\left( \boldsymbol{o}_{i-1}, \boldsymbol{c}_d \right)&		i=N_s+1, \cdots ,N_c\\
			\mathrm{B}_i\left( \boldsymbol{o}_{i-1} \right)&		i=N_c+1, \cdots ,N+1\\
			\mathrm{B}_{i}^{s}\left( \mathrm{CL}\left( \boldsymbol{o}_{i-1}, \boldsymbol{o}_{2N+2-i} \right) \right)&		i=N+2, \cdots ,2N+1\\
		\end{cases}
		\\
	\end{aligned}
\end{equation} 
where $\mathrm{B}_{i}$ denotes the $i$-th CDiT/DiT block, and $\boldsymbol{o}_i$ represents its corresponding output. The operation ``$\mathrm{CL}$'' refers to the concatenation followed by a linear projection. Finally, we discard the first token of the final output sequence $\boldsymbol{o}_{N+1}$, and apply a Layer Normalization, a linear projection, and an unpatchify layer to convert the remaining sequence back into a spatial predicted noise component $\bm{\hat{\epsilon}}_{t}$, which has the same shape as $\bm{z}_t$.

\subsubsection{CDiT Block Design}
We adopt a channel fusion method atop a standard Vision Transformer block \cite{Dosovitskiy2020AnII} to construct our CDiT block, as illustrated in Fig. \ref{fig_CDiT_network}. Unlike class labels or text prompts, our two conditional inputs are both fine-grained, sharing the same spatial dimensions as the diffusion hidden features $\boldsymbol{o}_{i}$. Compared to previous approaches such as cross-attention or in-context conditioning, channel fusion better preserves spatial consistency while incurring negligible computational overhead. Specifically, $\bm{c}_s$/$\bm{c}_d$ are first projected and extended into sequences $\boldsymbol{o}'\in \mathbb{R} ^{\left(L+1\right)\times D'}$ through an adapter. All CDiT blocks with the same conditions share the same adapter. $\boldsymbol{o}'$ is then concatenated with the input embeddings $\bm{o}_{i-1}\in \mathbb{R} ^{\left(L+1\right)\times D}$ along the channel dimension and subsequently passed through a linear layer to project the channel size back to $D$, resulting in fused embeddings $\boldsymbol{o}''$. The fused embeddings are finally processed by the standard DiT block to yield the output embeddings $\bm{o}_i$.

\subsubsection{Classifier-Free Guidance}
To further enhance the effectiveness of conditional guidance, we employ the widely-used classifier-free guidance (CFG) technique. In our setting, CFG is used to encourage the sampling process to generate samples $\bm{z}$ such that $\log p\left( \boldsymbol{z}|\boldsymbol{c}\right) $ is maximized, where $\bm{c}$ includes $\boldsymbol{c}_s$ and $\boldsymbol{c}_d$. By Bayes Rule, $\log p\left( \boldsymbol{c}|\boldsymbol{z} \right) \propto \log p\left( \boldsymbol{z}|\boldsymbol{c} \right) -\log p\left( \boldsymbol{z} \right)$ and hence $\nabla _{\boldsymbol{z}}\log p\left( \boldsymbol{c}|\boldsymbol{z} \right) \propto \nabla _{\boldsymbol{z}}\log p\left( \boldsymbol{z}|\boldsymbol{c} \right) -\nabla _{\boldsymbol{z}}\log p\left( \boldsymbol{z} \right) 
$. By interpreting the output of diffusion models as the score function, the DDPM sampling procedure can be guided by:
\begin{equation}\label{CFG}
	\begin{aligned}
		\boldsymbol{\hat{\epsilon}}_t=&\,\epsilon _{\theta}\left( \boldsymbol{z}_t,t,\boldsymbol{\phi },\boldsymbol{\phi } \right) \\
		&+w\left[ \epsilon _{\theta}\left( \boldsymbol{z}_t,t,\boldsymbol{c}_s,\boldsymbol{c}_d \right) -\epsilon _{\theta}\left( \boldsymbol{z}_t,t,\boldsymbol{\phi },\boldsymbol{\phi } \right) \right],
	\end{aligned}
\end{equation} 
where $w$ denotes the guidance scale, and $\boldsymbol{\phi}$ is a zero embedding to replace $\bm{c}_s$ and $\bm{c}_d$. During training, we randomly apply $\boldsymbol{\phi}$ with a certain probability $p_u$, enabling the model to learn to handle both cases. \add{In all reported experiments, the guidance scale is fixed to $w=2$; this value is selected through a grid search over candidate guidance scales and then kept unchanged for every channel setting.}

\subsection{KC-Inspired Bandwidth Allocation} \label{sectionIII_C}

Our dual-branch design introduces a fundamental challenge: how to allocate bandwidth between the semantic and detail branches under a total bandwidth constraint. Conventional approaches typically rely on entropy estimation via variational modeling to guide bandwidth allocation (BA), as adopted in rate-distortion (RD) optimized JSCC frameworks such as NTSCC \cite{dai2022nonlinear}. However, in DiT-JSCC, decoding is performed in a generative manner, where the transmitted signal serves as a conditioning input to guide posterior sampling, rather than supporting discriminative reconstruction as in conventional JSCC methods. As a result, DiT-JSCC operates under a rate-distortion-perception (RDP) optimization framework \cite{blau2019rethinking,zhang2021universal}, in which entropy-based BA tends to bias the optimization toward distortion minimization at the expense of perceptual and semantic fidelity. This mismatch motivates the need for a new BA strategy tailored for generative decoding.

\begin{figure}[t]
	\setlength{\abovecaptionskip}{0cm}
	\setlength{\belowcaptionskip}{0cm}
	\hspace{0cm}
	\centering{\includegraphics[scale=0.5]{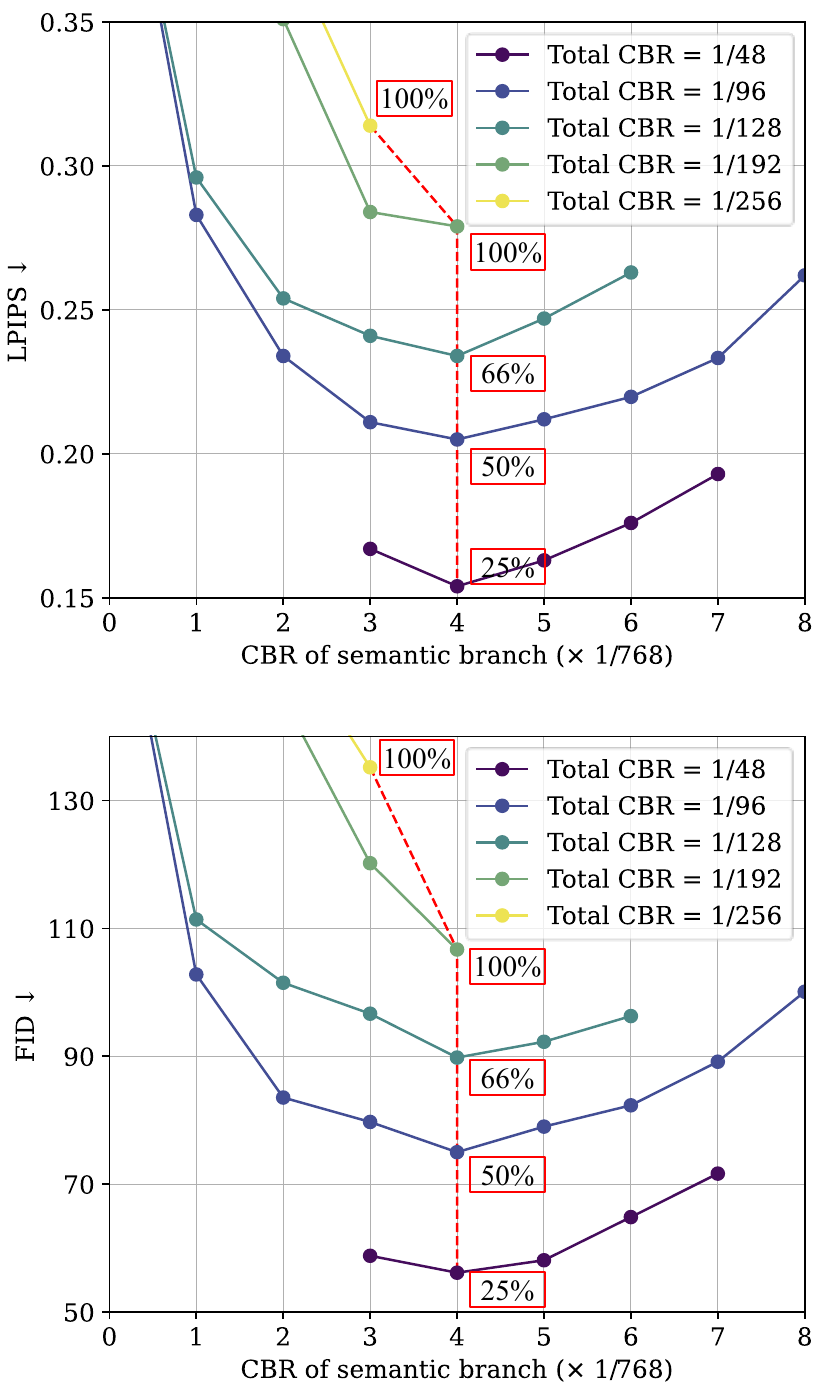}}%0.31
	\caption{Reconstruction quality (LPIPS and FID) under different bandwidths of the semantic branch with fixed-length encoding (without KC-inspired bandwidth allocation). All points on the same curve have the same total CBR $\rho$, with the x-axis indicating the CBR allocated to the semantic branch $\rho_s$. The red dashed lines connect the optimal allocation point for each total bandwidth budget, with the percentage of semantic bandwidth relative to the total bandwidth annotated at each point.}
	\label{fig_BA_analysis}
	\vspace{0em}
\end{figure}

To this end, we conduct exploratory experiments, as shown in Fig. \ref{fig_BA_analysis}, by exhaustively searching for near-optimal fixed bandwidth proportions between the semantic and detail branches. Across a wide range of total CBRs $\rho$, the optimal performance consistently corresponds to a similar semantic bandwidth level, as indicated by the red dashed line. When the semantic bandwidth $\rho_s$ falls below this saturation point, reconstruction quality degrades rapidly. In contrast, allocating additional bandwidth beyond this point to the semantic branch leads to only mild performance degradation due to reduced detail bandwidth. Notably, under extremely low total bandwidth, transmitting only the semantic branch yields the best overall performance. These observations confirm that the semantic branch plays a dominant role in ensuring generative reconstruction quality, while the detail branch provides secondary refinement.

\begin{figure}[t]
	\setlength{\abovecaptionskip}{0cm}
	\setlength{\belowcaptionskip}{0cm}
	\centering{\includegraphics[scale=0.56]{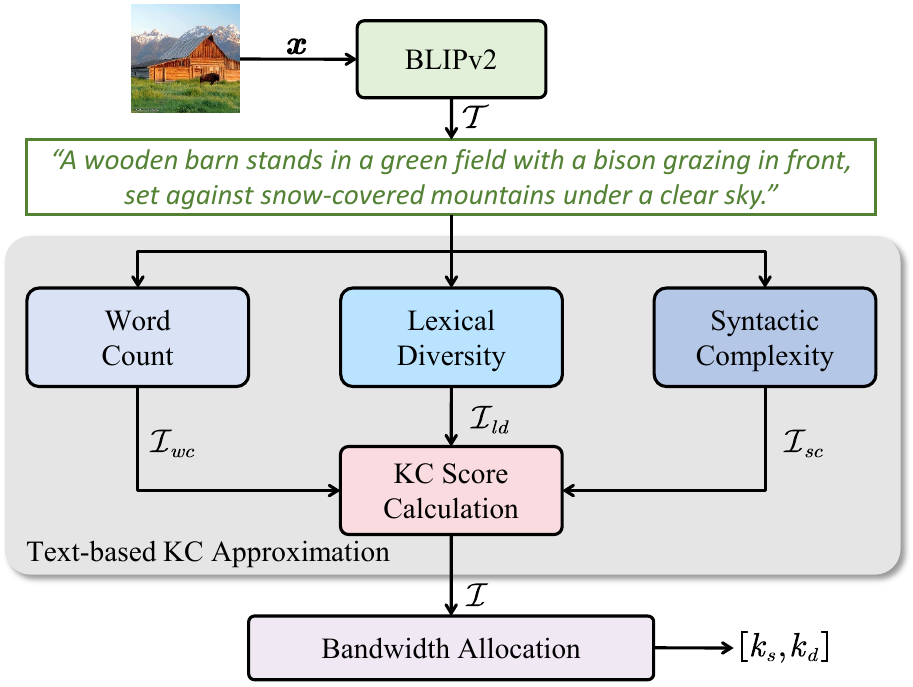}}
	\caption{The process of the proposed KC-inspired bandwidth allocation.}
	\label{fig_KC_BA_process}
	\vspace{0em}
\end{figure}

Beyond these empirical findings, we further argue that a fixed BA strategy is inherently suboptimal from the perspective of Kolmogorov complexity (KC) \cite{li2008introduction}, as it fails to account for instance-level variations in representation complexity and semantic familiarity. To address this limitation, we propose a KC-inspired instance-adaptive BA strategy. The core idea is to approximate Kolmogorov complexity through text-based semantic descriptions, thereby establishing a direct connection between BA and the semantic content of each visual sample. As illustrated in Fig. \ref{fig_KC_BA_process}, we first employ the BLIPv2 model \cite{li2023blip} to generate descriptive captions $\mathcal{T}$ that capture the core semantics of the input image $\bm{x}$. The complexity of these captions is then analyzed to guide sample-wise bandwidth reallocation. In contrast to traditional syntactic entropy estimation, which overemphasizes high-frequency patterns and low-level details, the proposed strategy more effectively reflects semantic richness and relevance. \add{Notably, BLIPv2 is used only for estimating semantic complexity in the transmitter-side BA module, rather than as a transmitted condition for generative decoding. Thus, possible captioning imperfections are reflected in the estimated semantic/detail rate split. Since captions are neither sent through the wireless channel nor required by the receiver, they do not introduce additional transmission-error risk.}

After obtaining the caption $\mathcal{T}$, we perform the following linguistic analyses:
\begin{itemize}
	\item \emph{Word Count (WC):} Let $\mathcal{I}_{wc}=\mathrm{WC}\left( \mathcal{T} \right)$ denotes the total number of words, serving as a basic indicator of caption length and descriptive completeness.
	
	\item \emph{Lexical Diversity (LD):} Defined as the ratio of unique words to total words: $\mathcal{I}_{ld} =\mathrm{UniqueWords}\left( \mathcal{T} \right)/\mathcal{I}_{wc}$. This metric captures the vocabulary richness of the caption and is known to correlate with the expressiveness of the description.
	
	\item \emph{Syntactic Complexity (SC):} To measure structural complexity, we parse $\mathcal{T}$ into sentences and compute the average number of tokens per sentence:
	\begin{equation}\label{SC}
		\begin{aligned}
			\mathcal{I}_{sc} =\frac{1}{N_{st}}\sum_{i=1}^{N_{st}}{|\mathrm{TokenNumber}\left( \bm{st}_i \right) |},
		\end{aligned}
	\end{equation}
	where $N_{st}$ is the number of sentences, and $\bm{st}_i$ denotes the $i$-th sentence. %Higher values of $\mathcal{I}_{sc}$ typically indicate more complex grammatical structures and richer syntactic patterns.
\end{itemize}

Then, we calculate a normalized KC score that jointly considers the above three metrics:
\begin{equation}\label{SC}
	\begin{aligned}
		&\mathcal{I} =\delta _{wc}\cdot \mathrm{Norm}\left( \mathcal{I} _{wc} \right) +\delta _{ld}\cdot \mathrm{Norm}\left(\mathcal{I} _{ld} \right) \\
		&\,\,\,\,\,\,\,\,\,\,+\delta _{sc}\cdot \mathrm{Norm}\left( \mathcal{I} _{sc} \right) ,
	\end{aligned}
\end{equation}
where $\mathrm{Norm}\left( \cdot \right)$ denotes the min-max normalization to normalize each component, and $\delta_{wc} ,\delta_{ld} ,\delta_{sc} \in \left[ 0,1 \right] $ are weighting coefficients satisfying $\delta_{wc} +\delta_{ld} +\delta_{sc} =1$. The normalization is performed across all captions in the evaluation set to ensure comparability.

\begin{algorithm}[t]
	%\footnotesize
	\caption{The inference process of DiT-JSCC}\label{alg1}
	\begin{algorithmic}[1] 
		\Require {Input image $\bm{x}$, time steps $T$, channel response $\bm{h}$, CFG factor $w$, network modules $E_{\mathrm{VFM}}$, $f_{\mathrm{PD}}$, $f_{\mathrm{LD}}$, $\mathcal{H}_s$, $\mathcal{H}_d$, $\bm{\epsilon} _{\theta}$, $\mathcal{D}$.}
		\LineComment{KC-inspired bandwidth allocation}
		\State $\mathcal{T} \gets \mathrm{BLIPv2}\left( \boldsymbol{x} \right)$
		\State $\mathcal{I} \gets \mathrm{KC\_Approximation}(\mathcal{T} )$
		\State $\left( k_s,k_d \right) \gets \mathrm{Bandwidth\_Allocation}(\mathcal{I} )$
		\LineComment{Dual-branch transmission}
		%\State $\boldsymbol{c}_s\xleftarrow{\mathcal{W} (\cdot ,\boldsymbol{h})}\boldsymbol{s}_s\xleftarrow{\mathcal{H}_s \left( f_{\mathrm{LD}}(\cdot ),k_s \right)}\boldsymbol{y}\xleftarrow{E_{\mathrm{VFM}}(\cdot )}\boldsymbol{x}$
		\State $\boldsymbol{y}\gets E_{\mathrm{VFM}}\left( \boldsymbol{x} \right) $
		\State $\boldsymbol{s}_s\gets \mathcal{H}_s \left( f_{\mathrm{LD}}\left( \boldsymbol{y} \right) ,k_s \right)$
		\State $\boldsymbol{s}_d\gets \mathcal{H}_d \left( f_{\mathrm{PD}}\left( \boldsymbol{x} \right) ,k_d \right) $
		\State $\boldsymbol{c}_s\gets \mathcal{W} \left( \boldsymbol{s}_s,\boldsymbol{h} \right) ,\boldsymbol{c}_d\gets \mathcal{W} \left( \boldsymbol{s}_d,\boldsymbol{h} \right) $
		\LineComment{Generative decoding}
		\State Initialize $\bm{z}_T \sim \mathcal{N} (\bm{0}, \mathbf{I})$
		\For {$t=T, \ldots , 1$}
		\LineComment{Classifier-free guidance}
		\State $\boldsymbol{\hat{\epsilon}}_t \gets \bm{\epsilon} _{\theta}\left( \boldsymbol{z}_t,t,\boldsymbol{\phi },\boldsymbol{\phi } \right) +w\left[ \bm{\epsilon} _{\theta}\left( \boldsymbol{z}_t,t,\boldsymbol{c}_s,\boldsymbol{c}_d \right) -\bm{\epsilon} _{\theta}\left( \boldsymbol{z}_t,t,\boldsymbol{\phi },\boldsymbol{\phi } \right) \right]$
		\State $\boldsymbol{\mu }_{\theta}\left( \boldsymbol{z}_{\boldsymbol{t}},t \right) \gets \frac{1}{\sqrt{\alpha _t}}\left( \boldsymbol{z}_t-\frac{1-\alpha _t}{\sqrt{1-\bar{\alpha}_t}}\boldsymbol{\hat{\epsilon}}_t \right) $
		\State Sample $\bm{z}_{t-1}$ from $\mathcal{N} \left( \boldsymbol{\mu }_{\theta}\left( \boldsymbol{z}_t,t \right) ,\sigma _t\mathbf{I} \right) $
		\EndFor
		\State $\boldsymbol{\hat{x}}\gets \mathcal{D} \left( \boldsymbol{z}_0 \right) $
		\State \textbf{return} $\bm{\hat{x}}$
	\end{algorithmic} 
\end{algorithm}

After that, we perform BA in an instance-adaptive manner based on KC score. Given a target transmission bandwidth, the total number of channel symbols $k$ can be determined, and the bandwidths of the two branches must satisfy $k_s+k_d=k,k_s\geqslant 0,k_d\geqslant 0$. Following Fig. \ref{fig_BA_analysis}, an optimal fixed allocation scheme is available, i.e., $k_s=\bar{k}_s, k_d=\bar{k}_d$. Building on this, we dynamically adjust the semantic branch bandwidth for each image, following the principle that images with higher KC score should be allocated more bandwidth to ensure semantic preservation. This can be formulated as 
\begin{equation}\label{SC_BA}
	\begin{aligned}
		&k_s=\bar{k}_s+\lceil \eta \cdot \left( \mathcal{I} -\bar{\mathcal{I}} \right) \cdot \bar{k}_s \rceil,\\
		&k_d=k-k_s,\\
	\end{aligned}
\end{equation}
where the bandwidth scaling factor $\eta$ maps the KC score to the corresponding channel bandwidth cost, $\lceil\cdot\rceil$ denotes the ceiling operation, and $\bar{\mathcal{I}}$ is the average KC score. We note that our method is plug-and-play during the inference stage and does not incur any additional training cost. \add{Because $k_s$ and $k_d$ take values from the candidate sets $\mathcal{K}_s$ and $\mathcal{K}_d$, the receiver only needs the corresponding candidate indices. The required side information is $\lceil\log_2 |\mathcal{K}_s|\rceil+\lceil\log_2 |\mathcal{K}_d|\rceil$ bits per image, where $|\cdot|$ denotes the cardinality of a candidate set. In our settings, this overhead is only 8 bits and can be appended as a compact header or transmitted through a reliable control link; it is negligible compared with the number of transmitted channel symbols.} The whole inference process of our DiT-JSCC is given in Algorithm \ref{alg1}.

\section{Experimental Results} \label{section_experimental_results}

\subsection{Experimental Setup}

\subsubsection{Datasets}
Our model is trained under the ImageNet train set \cite{deng2009imagenet}, which is widely used for evaluating class-conditional generative models. 
During training, each image is resized and center-cropped to either $256\times256$ or $512\times512$ resolution. 
For testing, following DiffCom \cite{wang2025diffcom}, we sample the same 100 images from the ImageNet validation set to construct a held-out subset.

\subsubsection{Metrics}
Our evaluation metrics consist of three pivotal dimensions: perceptual consistency, semantic consistency, and realism.
For \add{perceptual consistency}, we employ full-reference metrics including Learned Perceptual Image Patch Similarity (LPIPS) \cite{lpips} and Deep Image Structure and Texture Similarity (DISTS) \cite{dists}. These metrics focus on local structural and texture consistency within the feature space.
\add{For semantic consistency,} we incorporate CLIP \cite{radford2021learning}, Dreamsim \cite{NEURIPS2023_9f09f316}, and DINOv2 \cite{oquab2023dinov2} to evaluate high-level semantic alignment and conceptual fidelity.
For realism, we utilize the metric \add{Fr\'echet Inception Distance (FID)} \cite{fid} to quantify the distributional similarity between generated and real domains, ensuring visual authenticity and diversity. \add{For reporting convenience, score-inverted metrics marked with * are used only when a higher-is-better convention is desired; for example, LPIPS* and Dreamsim* indicate inverted versions of the original distance metrics.}

\subsubsection{Compared Methods}
We compare DiT-JSCC against a range of representative image transmission schemes. 
For separation-based schemes, we consider both engineered and learned image compression codecs for source coding, and combine them with 5G LDPC channel coding \cite{richardson2018design} for channel coding and digital modulation. 
After traversing the given combinations of LDPC coded modulation schemes, we use a 1/3 rate (4096, 6144) LDPC code with 4-ary quadrature amplitude modulation (4QAM) to ensure reliable transmission and the highest efficiency at SNR = $0$dB, a 1/2 rate (4096, 8192) LDPC + 4QAM at SNR = $2$dB, and a 1/4 rate (2048, 8192) LDPC + 4QAM at SNR = $-1$dB.

Specifically, the engineered image codecs include BPG \cite{BPG}, an HEVC-compliant intra-frame coding approach, and VTM \cite{bross2021developments}, the reference intra-frame codec for the VVC standard. The learned codecs include PerCo \cite{careil2023towards} and DiffEIC \cite{li2024towards}, which represent recent diffusion-driven advances in perceptual image compression.
For image JSCC schemes, we compare with SwinJSCC \cite{yang2024swinjscc} and two notable diffusion-based JSCC methods: DiffJSCC \cite{yang2024diffusion} and DiffCom \cite{wang2025diffcom}. 
Some channel-denoising-based GenJSCC methods \cite{wu2024cddm, zhang2025semantics} are excluded from certain comparisons, as their publicly available implementations are currently restricted to low-resolution inputs (e.g., $128\times128$).
Since DiffCom supports inference only at $256 \times 256$ resolution, whereas PerCo, DiffEIC, and DiffJSCC are designed for $512 \times 512$ or higher resolutions, we report all results in resolution-aligned groups across the evaluated schemes for fair comparison.

\subsubsection{Implementation Details}
For the visual foundation model $E_{\mathrm{VFM}}$, we adopt the ``dinov2-vit-b'' configuration of DINOv2 \cite{oquab2023dinov2}. The encoder structures $f_{\mathrm{PD}}$ and $f_{\mathrm{LD}}$ follow the SwinJSCC architecture: $f_{\mathrm{PD}}$ consists of four downsampling stages, while $f_{\mathrm{LD}}$ includes a single stage without downsampling. Our DiT decoder contains $N = 21$ DiT/CDiT blocks, with the first six designed as CDiT blocks. Each block uses 16 attention heads with a hidden dimension of $D = 1024$. Additionally, we incorporate a pre-trained VAE model from Stable Diffusion \cite{rombach2022high} to construct the latent space for diffusion.
To achieve flexible bandwidth control of signal branch, \add{we predefine two candidate sets for the semantic and detail branches, denoted by $\mathcal{K}_s$ and $\mathcal{K}_d$, respectively. During training, the number of channel symbols is randomly sampled as $k_s\in\mathcal{K}_s$ and $k_d\in\mathcal{K}_d$. Since each selected value can be represented by its index in the corresponding candidate set, $|\mathcal{K}_s|$ and $|\mathcal{K}_d|$ denote the numbers of available choices for the semantic and detail branches.}
Specifically, for the $512\times512$ model, \add{we set $\mathcal{K}_d=\left\{ 8i\times 256|i=0,1,\cdots ,12 \right\}$ and $\mathcal{K}_s=\left\{ 8i\times 256|i=2,\cdots ,14 \right\}$.}
For the $256\times256$ model, \add{we set $\mathcal{K}_d=\left\{2i\times 256|i=0,1,\cdots ,12 \right\}$ and $\mathcal{K}_s=\left\{ 2i\times 256|i=2,\cdots ,14 \right\}$.}
Note that we do not include $k_s = 0$ in training, as semantic signals are always considered essential regardless of bandwidth constraints. Regarding the wireless channel setting, we train separate models for each SNR under both AWGN and Rayleigh fading channels.

The training of DiT-JSCC follows the standard procedure of LDM. In our framework, the semantic branch (excluding $E_{\mathrm{VFM}}$), the detail branch, and the CDiT module are jointly optimized by the loss function \eqref{loss_DM}. At each training iteration, the discrete timestep $t$ is uniformly sampled from ${\left\{ 1,2,\cdots ,1000 \right\} }$, \add{and the CFG scale is fixed to $w=2$ for all reported experiments.} To accelerate training, we initialize our CDiT using a pre-trained class-conditional DiT model provided in \cite{bao2023all}. Training is conducted on four NVIDIA RTX 4090 GPUs. The batch size is set to 96 for $256 \times 256$ models and 32 for $512 \times 512$ models, with all models trained for 300,000 iterations at a fixed learning rate of $10^{-4}$.

\subsection{Experimental Results}

\begin{figure*}[t]
	\setlength{\abovecaptionskip}{0cm}
	\setlength{\belowcaptionskip}{0cm}
	\centering{\includegraphics[scale=0.45]{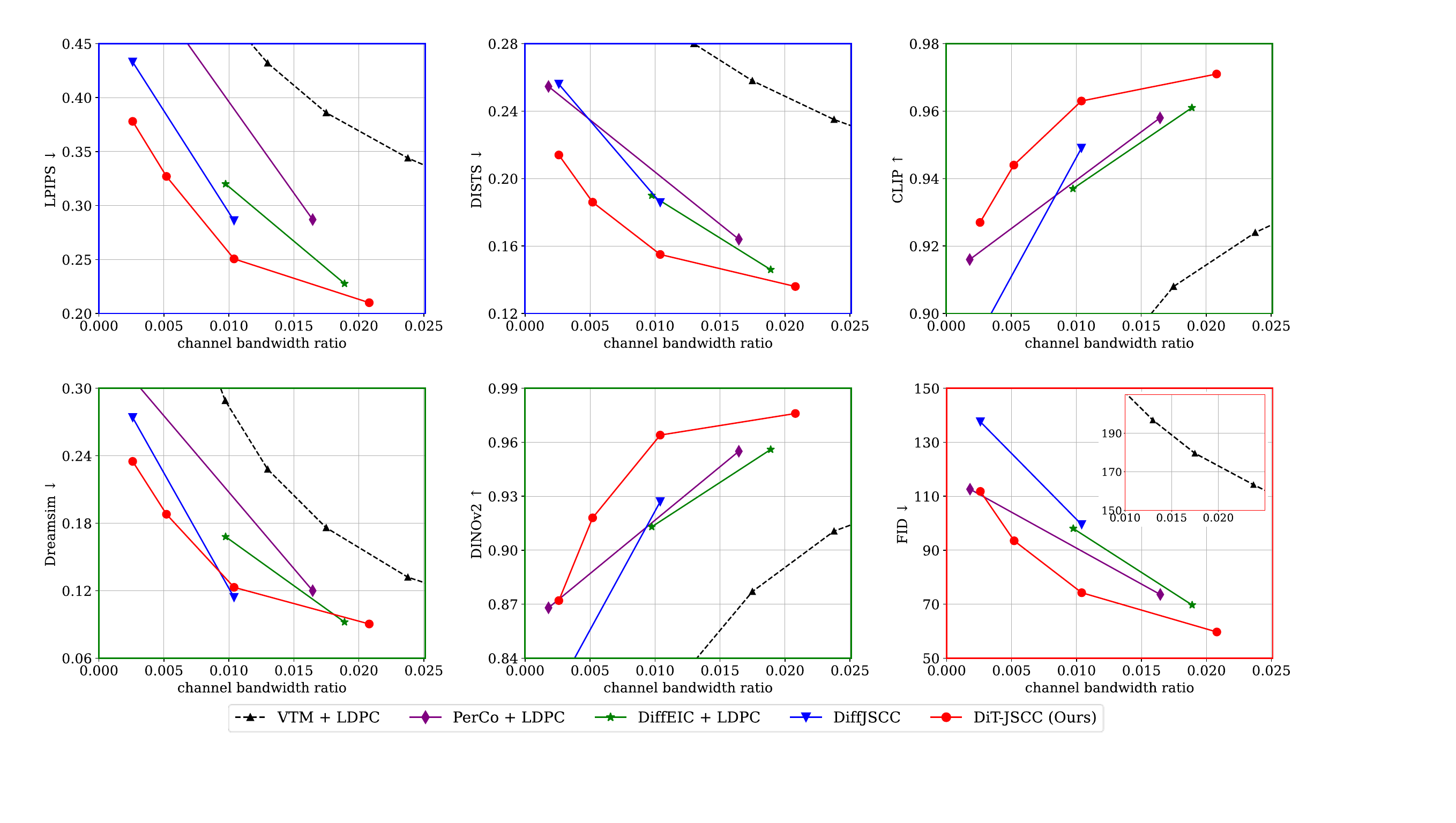}}
	\caption{Reconstruction performance versus channel bandwidth ratio, reported in consistency and realism metrics, tested on ImageNet $512\times512$ dataset, under AWGN channel (SNR = 0dB). Metrics of different categories are highlighted with different colored frames, where \add{blue frames represent perceptual consistency metrics}, green frames represent higher-level semantic consistency, and red frame represents realism. The upward arrow ``$\uparrow$'' indicates that higher values of the metric are favorable, and vice versa. Inset plots are provided for selected metrics to compare all methods clearly.}
	\label{fig_results_CBR_metrics}
	\vspace{0em}
\end{figure*}

\begin{figure*}[t]
	\setlength{\abovecaptionskip}{0cm}
	\setlength{\belowcaptionskip}{0cm}
	\centering{\includegraphics[scale=0.45]{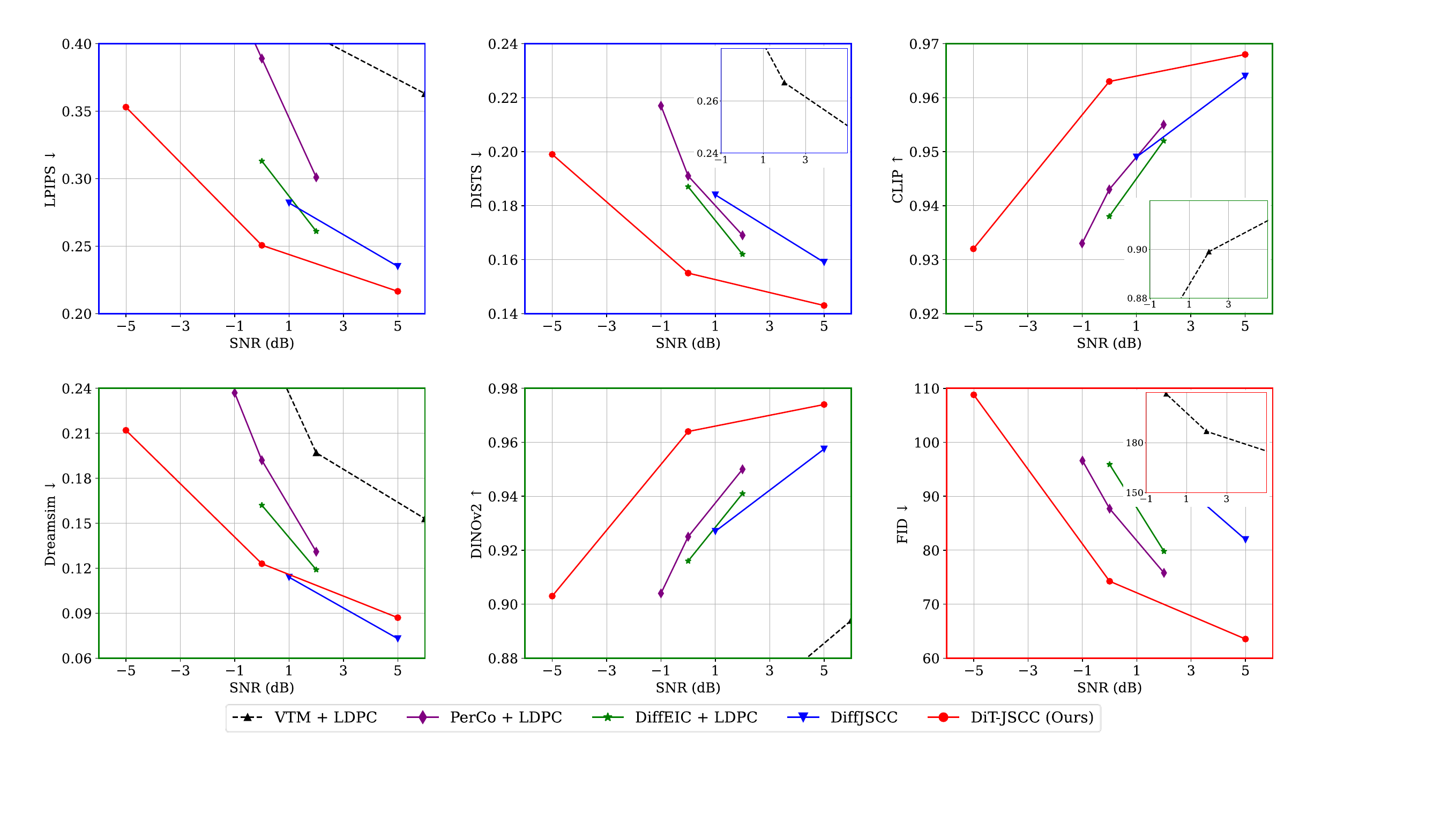}}
	\caption{Reconstruction performance versus SNR, reported in terms of consistency and realism metrics, tested on the ImageNet $512\times512$ dataset under the AWGN channel with a CBR of approximately 1/96.}
	\label{fig_results_SNR_metrics}
	\vspace{0em}
\end{figure*}

In this paper, we focus on the extremely low-bandwidth regime, where the CBR ranges from 1/384 to 1/48, and evaluate our method under both AWGN and Rayleigh fading channels across a low-SNR range of $-5$ dB to $5$ dB. \add{This scope reflects the main target of DiT-JSCC: robust semantic reconstruction when both channel budget and channel quality are severely constrained. As CBR increases, pixel-oriented baselines are expected to recover more low-level details, while the advantage of DiT-JSCC mainly comes from preserving semantic consistency in the bandwidth-limited region.}

Figure \ref{fig_results_CBR_metrics} reports the results of reconstruction performance versus CBR over the AWGN channel at SNR $=0$ dB. 
Among all compared methods, DiT-JSCC achieves substantially better performance across all consistency metrics while also attaining the best FID scores. 
These results confirm that our approach preserves content fidelity effectively  across diverse representation levels  without compromising realism.
Fig. \ref{fig_results_SNR_metrics} reports performance under varying channel SNR at a fixed CBR of 1/96. 
For separation-based methods, we report the lowest-SNR points they can achieve by progressively decreasing the channel-coding rate and modulation order, until the remaining CBR becomes insufficient for source coding.
It can be observed that the proposed DiT-JSCC consistently outperforms all baselines across all metrics, particularly in the lower channel SNR regime. 

Table \ref{tab:imagenet256_results} and Table \ref{tab:imagenet256_results_rayleigh} compare the channel-noise robustness of the proposed DiT-JSCC against existing JSCC methods under AWGN and Rayleigh fading channels, respectively.
Results show that DiT-JSCC consistently outperforms DiffCom and other JSCC schemes across all evaluation metrics, demonstrating superior robustness to channel variations. \add{For the Rayleigh-fading comparison, we further include HiFi-DiffCom with SwinJSCC as its JSCC backbone as a generative baseline. Since the open-source SwinJSCC models support SNRs of 1, 4, 7, 10, and 13 dB, we report HiFi-DiffCom using the nearest available 1 dB and 7 dB backbones for the 0 dB and 5 dB comparison settings, respectively.}

\begin{table*}[t] 
	\caption{Comparisons on ImageNet $256\times256$ with a CBR of 1/48 under AWGN channel.}  
	\vspace{0.3em}
	\setlength{\tabcolsep}{3pt}
	\centering
	\normalsize
	\begin{tabularx}{\textwidth}{@{}>{\hsize=0.50\hsize\centering\arraybackslash}X>{\hsize=2.35\hsize\centering\arraybackslash}X*{3}{>{\hsize=0.76\hsize\centering\arraybackslash}X}>{\hsize=1.08\hsize\centering\arraybackslash}X>{\hsize=0.76\hsize\centering\arraybackslash}X>{\hsize=1.03\hsize\centering\arraybackslash}X@{}}
		\toprule
		\multicolumn{2}{c}{ImageNet} & \multicolumn{2}{c}{\add{Perceptual Consistency}} & \multicolumn{3}{c}{\add{Semantic Consistency}} & Realism \\
		\cmidrule(lr){1-2}\cmidrule(lr){3-4}\cmidrule(lr){5-7}\cmidrule(lr){8-8}
		
		{SNR} & Method & LPIPS $\downarrow$  & DISTS $\downarrow$ & CLIP $\uparrow$ & Dreamsim $\downarrow$ & DINOv2 $\uparrow$ & FID $\downarrow$\\
		% 分组行：Consistency 覆盖5列（LPIPS~DINOv2），FID ��?Realism
		\midrule
		\multirow{6}{*}{0dB} 
		& \mbox{BPG + LDPC} & 0.444 & 0.314 & 0.828 & 0.402 & 0.717 & 252 \\
		& \mbox{VTM + LDPC} & 0.427 & 0.313 & 0.844 & 0.360 & 0.720 & 244 \\
		& \mbox{SwinJSCC (1dB)} & 0.261 & 0.233 & 0.915 & 0.178 & 0.878 & 138 \\
		& \mbox{HiFi-DiffCom (Deep JSCC)} & 0.283 & 0.203 & 0.933 & 0.192 & 0.880 & 115 \\
		& \mbox{HiFi-DiffCom (NTSCC)} & 0.211 & 0.191 & 0.936 & 0.143 & 0.900 & 107 \\
		& \cellcolor{gray!12}DiT-JSCC (Ours) & \cellcolor{gray!12}\textbf{0.166} & \cellcolor{gray!12}\textbf{0.151} & \cellcolor{gray!12}\textbf{0.963} & \cellcolor{gray!12}\textbf{0.097} & \cellcolor{gray!12}\textbf{0.958} & \cellcolor{gray!12}\textbf{60} \\
		\midrule
		\multirow{6}{*}{5dB} 
		& \mbox{BPG + LDPC} & 0.351 & 0.272 & 0.874 & 0.271 & 0.808 & 197 \\
		& \mbox{VTM + LDPC} & 0.321 & 0.267 & 0.881 & 0.231 & 0.797 & 198 \\
		& \mbox{SwinJSCC (7dB)} & 0.188 & 0.194 & 0.939 & 0.104 & 0.937 &95 \\
		& \mbox{HiFi-DiffCom (Deep JSCC)} & 0.165 & 0.164 & 0.959 & 0.089 & 0.938 & 69 \\
		& \mbox{HiFi-DiffCom (NTSCC)} & 0.137 & 0.140 & 0.961 & 0.074 & 0.947 & 57 \\
		& \cellcolor{gray!12}DiT-JSCC (Ours) & \cellcolor{gray!12}\textbf{0.122} & \cellcolor{gray!12}\textbf{0.135} & \cellcolor{gray!12}\textbf{0.97} & \cellcolor{gray!12}\textbf{0.064} & \cellcolor{gray!12}\textbf{0.969} & \cellcolor{gray!12}\textbf{49} \\
		\bottomrule
	\end{tabularx}
	\label{tab:imagenet256_results}
	\vspace{-0.0cm}
\end{table*}

\begin{table*}[t] 
	\caption{Comparisons on ImageNet $256\times256$ with a CBR of 1/48 under Rayleigh fading channel.}  
	\setlength{\tabcolsep}{3pt}
	\centering
	\normalsize
	\begin{tabularx}{\textwidth}{@{}>{\hsize=0.50\hsize\centering\arraybackslash}X>{\hsize=2.15\hsize\centering\arraybackslash}X*{3}{>{\hsize=0.76\hsize\centering\arraybackslash}X}>{\hsize=1.08\hsize\centering\arraybackslash}X>{\hsize=0.76\hsize\centering\arraybackslash}X>{\hsize=1.23\hsize\centering\arraybackslash}X@{}}
		\toprule
		\multicolumn{2}{c}{ImageNet} & \multicolumn{2}{c}{\add{Perceptual Consistency}} & \multicolumn{3}{c}{\add{Semantic Consistency}} & Realism \\
		\cmidrule(lr){1-2}\cmidrule(lr){3-4}\cmidrule(lr){5-7}\cmidrule(lr){8-8}
		
		{SNR} & Method & LPIPS $\downarrow$  & DISTS $\downarrow$ & CLIP $\uparrow$ & Dreamsim $\downarrow$ & DINOv2 $\uparrow$ & FID $\downarrow$\\
		% 分组行：Consistency 覆盖5列（LPIPS~DINOv2），FID ��?Realism
		\midrule
		\multirow{5}{*}{0dB} 
		& BPG + LDPC & \multicolumn{6}{c}{\emph{The available bandwidth is insufficient to support BPG encoding}} \\
		& \mbox{VTM + LDPC} & 0.490 & 0.338 & 0.824 & 0.439 & 0.685 & 276 \\
		& \mbox{SwinJSCC (1dB)} & 0.297 & 0.252 & 0.897 & 0.236 & 0.833 & 169 \\
		& \add{\mbox{HiFi-DiffCom (SwinJSCC)}} & \add{0.249} & \add{0.216} & \add{0.924} & \add{0.181} & \add{0.883} & \add{126} \\
		& \cellcolor{gray!12}DiT-JSCC (Ours) & \cellcolor{gray!12}\textbf{0.185} & \cellcolor{gray!12}\textbf{0.173} & \cellcolor{gray!12}\textbf{0.943} & \cellcolor{gray!12}\textbf{0.121} & \cellcolor{gray!12}\textbf{0.934} & \cellcolor{gray!12}\textbf{77} \\
		\midrule
		\multirow{5}{*}{5dB} 
		& \mbox{BPG + LDPC} & 0.351 & 0.272 & 0.874 & 0.271 & 0.808 & 197 \\
		& \mbox{VTM + LDPC} & 0.321 & 0.267 & 0.881 & 0.231 & 0.797 & 198 \\
		& \mbox{SwinJSCC (7dB)} & 0.239 & 0.226 & 0.922 & 0.155 & 0.921 &128 \\
		& \add{\mbox{HiFi-DiffCom (SwinJSCC)}} & \add{0.218} & \add{0.199} & \add{0.944} & \add{0.136} & \add{0.916} & \add{100} \\
		& \cellcolor{gray!12}DiT-JSCC (Ours) & \cellcolor{gray!12}\textbf{0.143} & \cellcolor{gray!12}\textbf{0.151} & \cellcolor{gray!12}\textbf{0.959} & \cellcolor{gray!12}\textbf{0.083} & \cellcolor{gray!12}\textbf{0.952} & \cellcolor{gray!12}\textbf{63} \\
		\bottomrule
	\end{tabularx}
	\add{\footnotesize{Note: For HiFi-DiffCom (SwinJSCC), the 0 dB and 5 dB rows use the nearest available SwinJSCC backbones trained at 1 dB and 7 dB, respectively.}}
	\label{tab:imagenet256_results_rayleigh}
	\vspace{-0.0cm}
\end{table*}

\begin{figure*}[t]
	\setlength{\abovecaptionskip}{0cm}
	\setlength{\belowcaptionskip}{0cm}
	\centering{\includegraphics[scale=0.49]{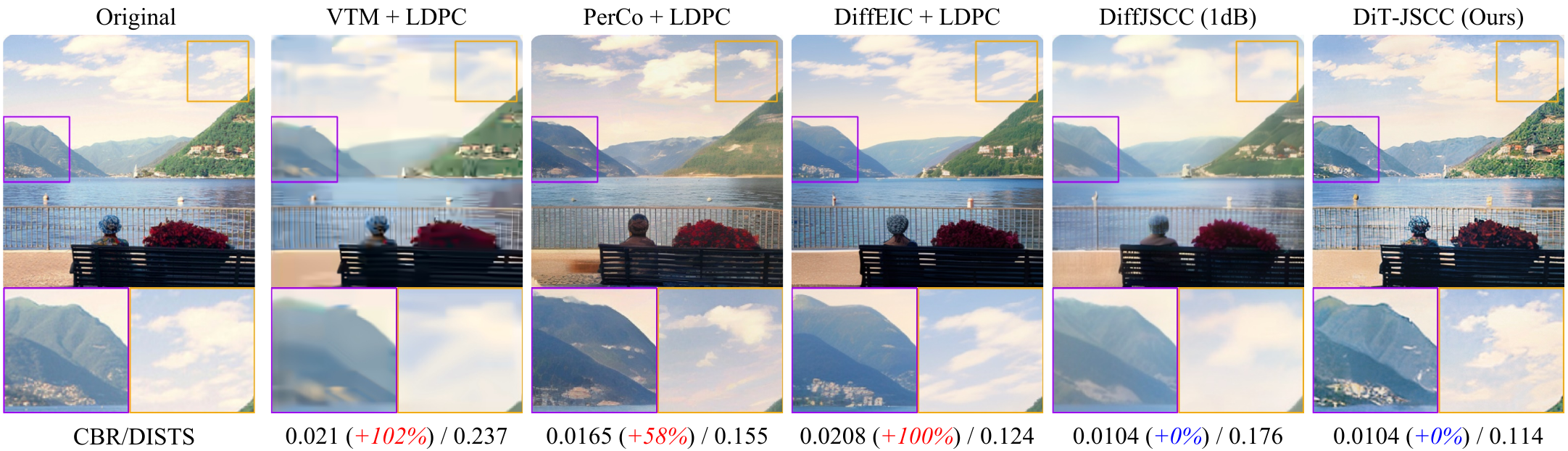}}
	\caption{Visual comparison on the ImageNet dataset $512\times512$ at 0dB SNR (except for DiffJSCC, which is 1dB) under the AWGN channel. For each image, we present two crops of it for detail comparison. Red numbers indicate the percentage of bandwidth cost increase compared to our DiT-JSCC. Note that our models always have the smaller CBR or lower SNR.}
	\label{fig_visual_results_512}
	\vspace{0em}
\end{figure*}

Figure \ref{fig_visual_results_512} gives qualitative comparisons on the ImageNet dataset. For reference, we also annotate each reconstruction with its CBR and DISTS score.
Among all these methods, ``VTM + LDPC'' exhibits the worst visual quality, producing noticeable blurry artifacts. 
In contrast, generative methods, including Perco, DiffEIC, DiffJSCC, and our DiT-JSCC, can generate reconstructions with realistic textures. 
However, as shown in Fig. \ref{fig_visual_results_512}, only DiT-JSCC accurately reconstructs the ridges and valleys of the mountain, faithfully preserving the underlying structure, while other methods exhibit noticeable \add{deviation} from the original image in low-level semantic elements, such as local geometry and structural details.

\begin{figure}[!t]
	\setlength{\abovecaptionskip}{0cm}
	\setlength{\belowcaptionskip}{0cm}
	\centering{\includegraphics[scale=0.31]{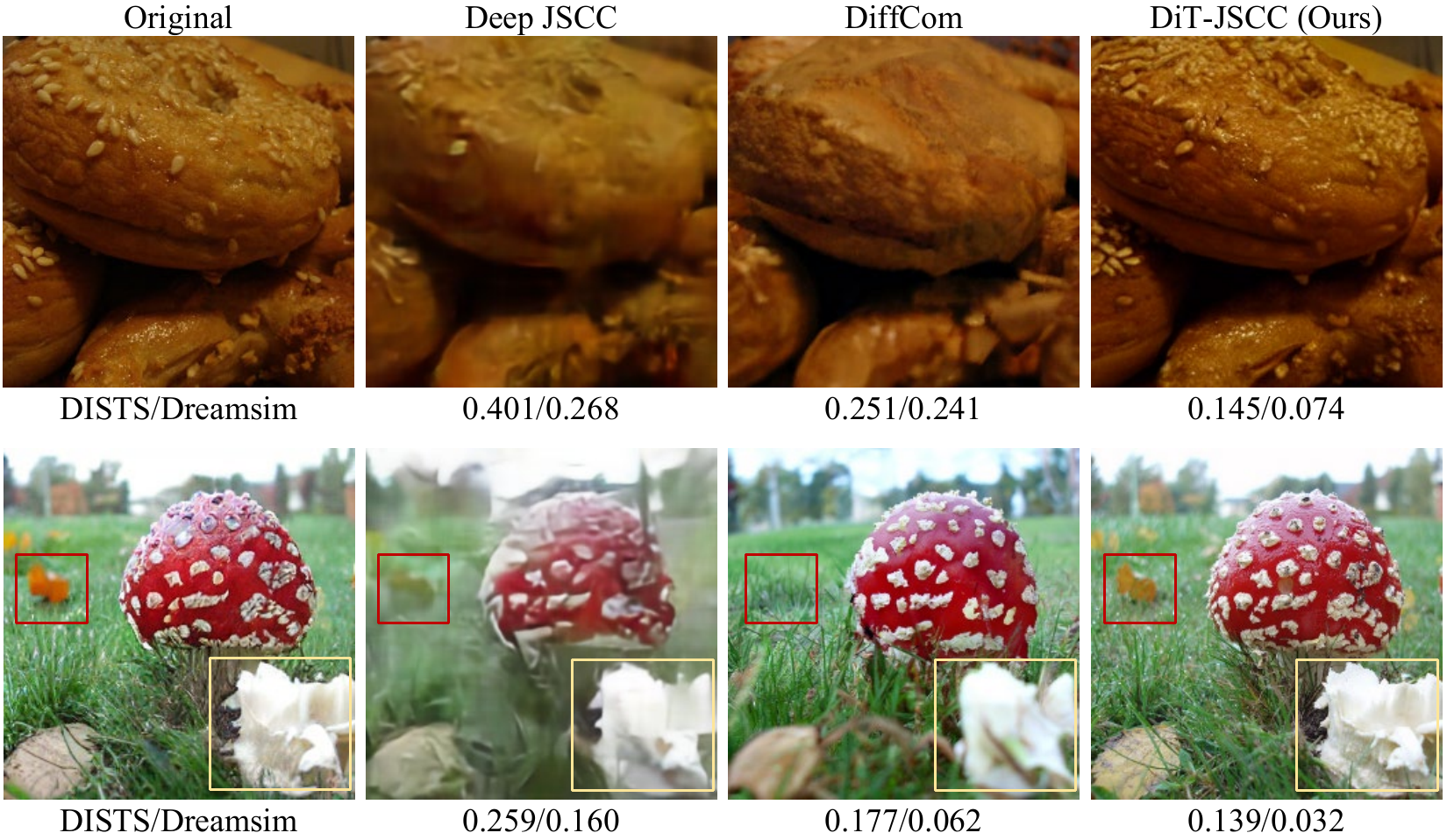}}
	\caption{Visual samples are provided on the ImageNet $256\times256$ dataset to demonstrate high-level semantic distortion. All reconstructions are obtained in an AWGN channel with SNR 0 dB and with the same CBR value of 1/48.}
	\label{fig_visual_results_256}
	\vspace{0em}
\end{figure}

To better illustrate this, Fig. \ref{fig_visual_results_256} presents additional visual results focusing on high-level semantic consistency.
When decoding from the same transmitted signal $\bm{s}$, it can be seen that DiffCom produces reconstructions in which primary objects are semantically distorted and hard to recognize, even more severely than in Deep JSCC.
We attribute this behavior to the limited high-level semantic content carried by $\bm{s}$, which can mislead the generative decoder and steer the diffusion sampling process toward incorrect reverse trajectories, resulting in large semantic deviations.
In contrast, our method explicitly incorporates high-level semantic information from the semantic branch, which guides the diffusion decoding process to produce semantically faithful reconstructions.
These results support our hypothesis that incorporating high-level semantic information into the diffusion decoding process is crucial for achieving faithful image reconstruction in GenJSCC systems.

\add{To further validate this observation from a caption-level semantic perspective, we conduct a small-scale LLM-based consistency evaluation on 100 aligned ImageNet samples. Following the description-comparison procedure in Fig. \ref{fig_sem_metrics}, the large language model GPT-5.5 first generates one textual caption for each original image and each reconstructed image under the same prompt. We then use the original-image caption as the reference and the reconstructed-image caption as the candidate, and compute BLEU-4 \cite{papineni2002bleu} and BERTScore-F1 \cite{zhang2020bertscore}. As shown in Table \ref{tab:llm_eval}, DiT-JSCC achieves higher caption consistency than VTM + LDPC, SwinJSCC, and DiffCom, indicating that its reconstructions better preserve the high-level semantic content of the original images.}

\begin{table}[!t]
	\caption{\add{Small-scale caption-based LLM semantic consistency evaluation.}}
	\setlength{\tabcolsep}{3pt}
	\centering
	\normalsize
	\begin{threeparttable}
	\begin{tabularx}{\columnwidth}{@{}>{\hsize=1\hsize\centering\arraybackslash}X>{\hsize=1\hsize\centering\arraybackslash}X>{\hsize=1\hsize\centering\arraybackslash}X@{}}
		\toprule[1.2pt]
		\multicolumn{3}{c}{\makecell{\textbf{Caption-based semantic} \textbf{consistency evaluation}}} \\
		\midrule[0.8pt]
		Method & \makecell{BLEU-4 (\%) $\uparrow$} & \makecell{BERTScore F1 $\uparrow$} \\
		\midrule[0.8pt]
		\mbox{VTM + LDPC} & 46.8 & 0.843 \\
		\mbox{SwinJSCC} & 61.7 & 0.920 \\
		DiffCom & 72.5 & 0.978 \\
		\rowcolor{gray!12}
		DiT-JSCC (Ours) & \textbf{96.3} & \textbf{0.995} \\
		\bottomrule[1.2pt]
	\end{tabularx}
	\begin{tablenotes}
		\footnotesize
		\item The large language model GPT-5.5 is used to generate captions for 100 aligned ImageNet samples. Metrics are computed between original-image captions and reconstructed-image captions.
	\end{tablenotes}
	\end{threeparttable}
	\label{tab:llm_eval}
	\vspace{-0.0cm}
\end{table}

\add{Finally, Table \ref{tab:complexity_latency} summarizes the parameter counts and encoding/decoding latency of representative diffusion-based methods. All listed methods use 50 diffusion steps, and their parameter counts are reported as Codec + Diffusion model where applicable. The table shows that the compared methods follow different complexity-latency profiles: DiffJSCC has the lowest encoding latency but the longest decoding time, DiffEIC + LDPC and PerCo + LDPC require larger codec-plus-diffusion-model parameter budgets with several-second decoding latency, and DiT-JSCC uses a smaller reported parameter count with intermediate encoding and decoding time among these diffusion-based methods. These results indicate that diffusion-based JSCC performance should be interpreted together with model size and inference cost.}

\begin{table}[!t]
	\caption{Comparison of parameter counts and encoding/decoding time among diffusion-based methods.}
	\setlength{\tabcolsep}{2pt}
	\centering
	\normalsize
	\begin{threeparttable}
	\begin{tabularx}{\columnwidth}{@{}>{\hsize=1.8\hsize\centering\arraybackslash}X>{\hsize=0.8\hsize\centering\arraybackslash}X>{\hsize=0.7\hsize\centering\arraybackslash}X>{\hsize=0.85\hsize\centering\arraybackslash}X>{\hsize=0.85\hsize\centering\arraybackslash}X@{}}
		\toprule[1.2pt]
		\multicolumn{5}{c}{\textbf{Model complexity and inference latency}} \\
		\midrule[0.8pt]
		Method & \makecell{Params\tnote{a}\\(M)} & \makecell{Diff.\\Steps} & \makecell{Enc.\\Time(ms)} & \makecell{Dec.\\Time(ms)} \\
		\midrule[0.8pt]
		DiffJSCC & 480+865 & 50 & 4.6 & 6711 \\
		DiffEIC + LDPC  & 514+865 & 50 & 153 & 4647 \\
		PerCo + LDPC  & 430+865 & 50 & 404 & 2817 \\
		\rowcolor{gray!12}
		DiT-JSCC (Ours) & 148+321 & 50 & 43 & 5301 \\
		\bottomrule[1.2pt]
	\end{tabularx}
	\begin{tablenotes}
		\footnotesize
		\item[a] Params reported as Codec + Diffusion model.
	\end{tablenotes}
	\end{threeparttable}
	\label{tab:complexity_latency}
	\vspace{-0.0cm}
\end{table}

\subsection{Ablation Study}

In this subsection, we conduct a series of ablation studies to validate the effectiveness of our key architectural choices. All experiments are performed over an AWGN channel with an SNR of 0 dB on the ImageNet $256\times256$ dataset.

\subsubsection{\texorpdfstring{Contributions of the Conditional Injection Strategy}{Contributions of the Conditional Injection Strategy}}
\add{To verify the conditional injection design introduced in Section \ref{section_method}, we conduct an ablation study on the CDiT conditioning strategy. Table \ref{tab:injection_ablation} compares the proposed hierarchical injection strategy with two representative alternatives. In uniform injection, both semantic and detail conditions are provided to all conditional blocks, whereas swapped injection reverses the semantic/detail injection order. The proposed strategy obtains the lowest LPIPS and DISTS, demonstrating the effectiveness of the CDiT conditioning design described in Section \ref{section_method}.}

\begin{table}[!t]
	\caption{Ablation on the conditional injection strategy in CDiT. Lower LPIPS and DISTS indicate better \add{perceptual reconstruction consistency}.}
	\setlength{\tabcolsep}{3pt}
	\centering
	\normalsize
	\begin{tabularx}{\columnwidth}{@{}>{\hsize=1.6\hsize\centering\arraybackslash}X>{\hsize=0.7\hsize\centering\arraybackslash}X>{\hsize=0.7\hsize\centering\arraybackslash}X@{}}
		\toprule[1.2pt]
		\multicolumn{3}{c}{\textbf{Conditional injection strategy in CDiT}} \\
		\midrule[0.8pt]
		Injection strategy & LPIPS $\downarrow$ & DISTS $\downarrow$ \\
		\midrule[0.8pt]
		Uniform injection & 0.191 & 0.1764 \\
		Swapped injection & 0.207 & 0.186 \\
		\rowcolor{gray!12}
		Hierarchical injection (Ours) & \textbf{0.166} & \textbf{0.151} \\
		\bottomrule[1.2pt]
	\end{tabularx}
	\label{tab:injection_ablation}
	\vspace{-0.0cm}
\end{table}

\subsubsection{Contributions of the Two Branch Design}
\begin{figure}[!t]
	\begin{center}
		\hspace{-.20in}
		\subfigure{\includegraphics[height=0.215\textwidth]{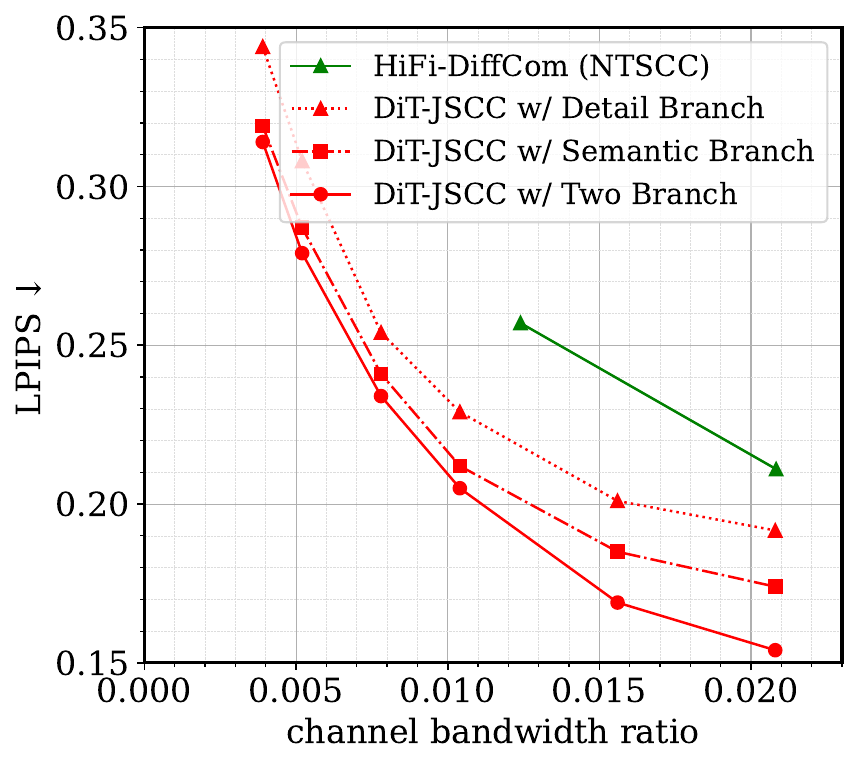}}
		\subfigure {\includegraphics[height=0.215\textwidth]{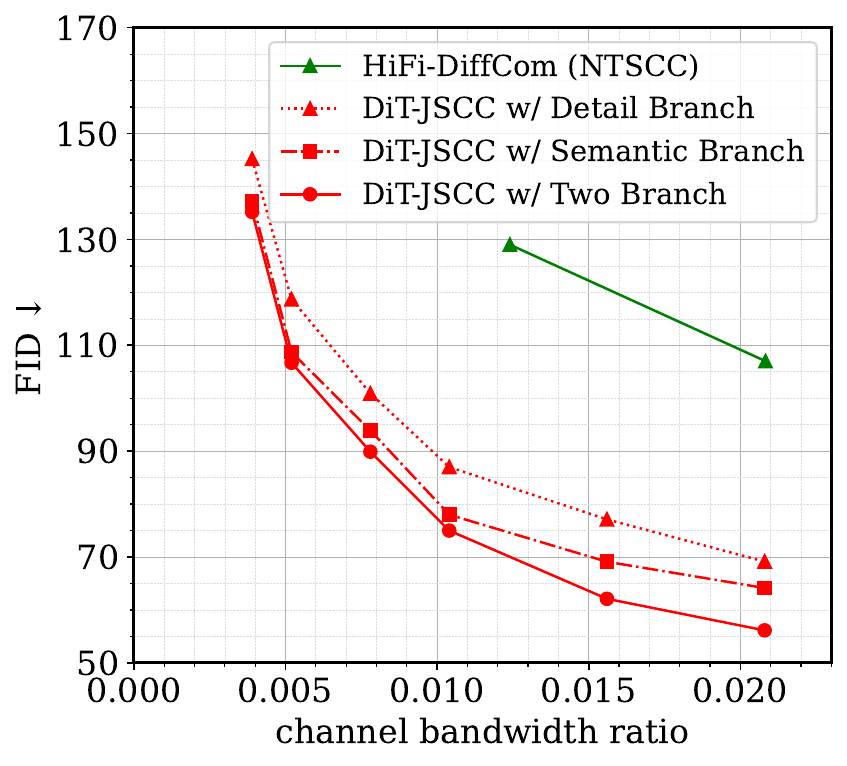}}
		\hspace{-.20in}
		\caption{Effect of different transmitter architectures and comparison with the HiFi-DiffCom (NTSCC) baseline, including the proposed dual branches, single detail branch, and single semantic branch.}
		\label{fig_ab_dual_branches}
	\end{center}
	\vspace{0em}
\end{figure}

To evaluate the impact of the proposed dual-branch architecture, we design two kinds of \emph{single-branch} variants of DiT-JSCC: one retaining only the detail branch (i.e., DiT-JSCC w/ detail branch), and the other retaining only the semantic branch (DiT-JSCC w/ semantic branch). 
Fig. \ref{fig_ab_dual_branches} presents the performance comparison among the three transmitter architectures, where all of them are trained under the same experimental settings without using the KC-inspired BA strategy. \add{We also include HiFi-DiffCom (NTSCC) as a baseline in the same figure to provide direct context for the ablation results.}
Clearly, DiT-JSCC w/ detail branch performs worse than the full model, due to the limited semantic representation capability of the pixel-domain JSCC encoder learned from scratch. 
In contrast, DiT-JSCC w/ semantic branch achieves comparable performance to the full model under low CBR conditions but lags behind as the CBR increases. 
This behavior can be explained by the limited capacity of VFM features to capture fine-grained details.
By integrating both branches, the proposed DiT-JSCC achieves consistently superior performance across various bandwidth regimes, which validates that the semantic and detail branches are complementary to each other.
\add{This trend also helps relate the ablation to the baseline comparisons in Figs. \ref{fig_results_CBR_metrics}--\ref{fig_results_SNR_metrics} and Tables \ref{tab:imagenet256_results}--\ref{tab:imagenet256_results_rayleigh}. Compared with HiFi-DiffCom (NTSCC), the single-branch DiT-JSCC variants still benefit from the proposed conditional generative decoding paradigm, indicating that the decoder-side paradigm itself is an important source of performance gain. The semantic-only branch is already strong in the extremely low-CBR regime because high-level semantic guidance is the main bottleneck for generative decoding. However, as more bandwidth becomes available, the full dual-branch model becomes more advantageous because the detail branch contributes complementary texture and local structure. The detail-only variant remains weaker because low-level detail features alone cannot provide reliable semantic guidance to the DiT decoder. Therefore, the final performance gain should be attributed to both the conditional generative decoding paradigm and the complementary dual-branch transmitter design, rather than to a simple bandwidth hyperparameter choice.}

\subsubsection{Contributions of the KC-Inspired Bandwidth Allocation Strategy}
\begin{figure}[!t]
	\setlength{\abovecaptionskip}{0cm}
	\setlength{\belowcaptionskip}{0cm}
	\centering{\includegraphics[scale=0.33]{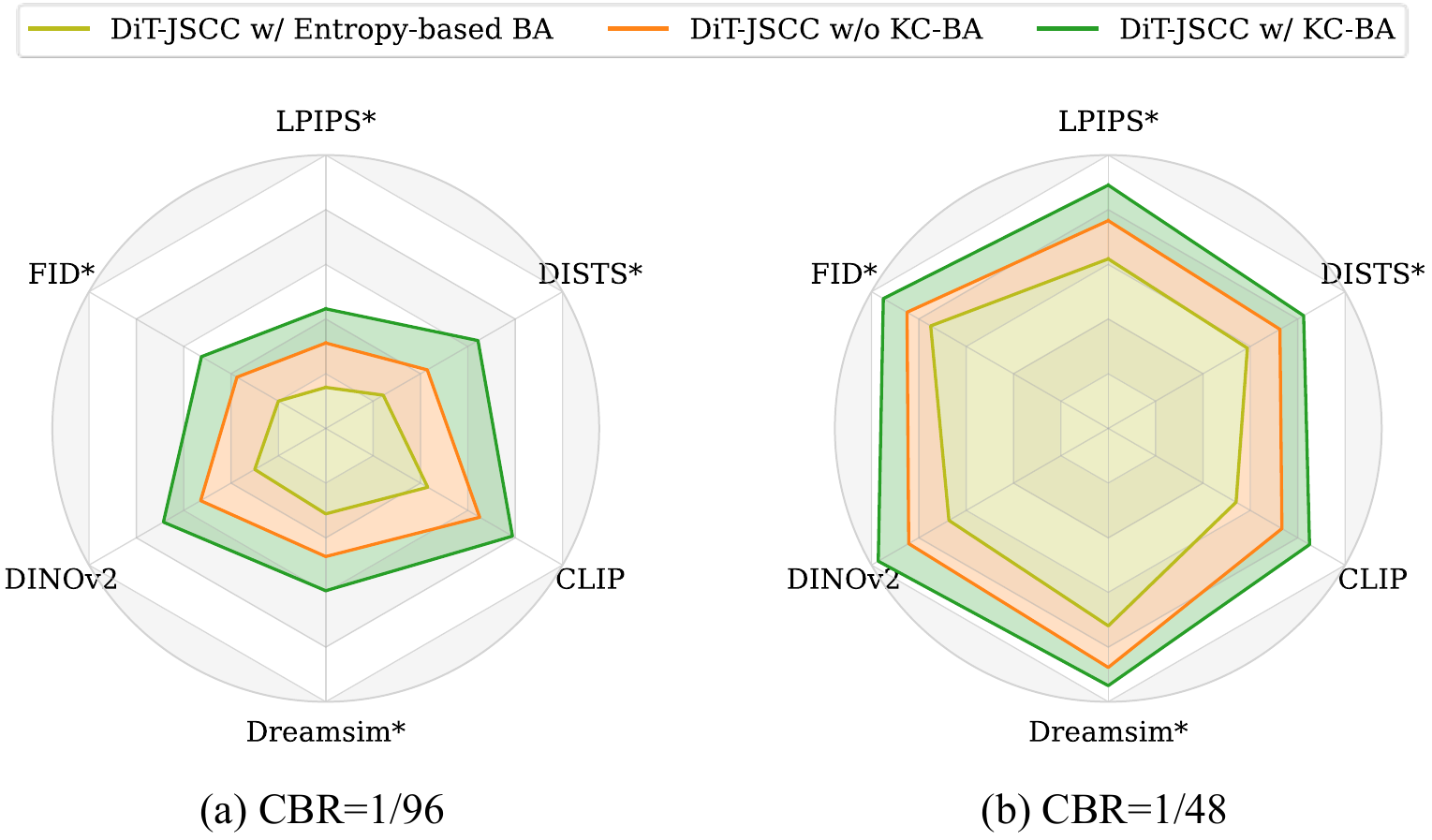}}
	\caption{Impact of different bandwidth allocation strategies. The radar charts compare six metrics under two CBR settings. * denotes the use of a score inversion operation for better comparison, e.g., $\text{LPIPS*} = 1 - \text{LPIPS}$. In two subfigures, the axes corresponding to each metric use the same value range.}
	\label{fig_ab_KC_BA}
	\vspace{0em}
\end{figure}

The BA strategy plays a vital role in DiT-JSCC, as analyzed before in Fig. \ref{fig_BA_analysis}.
To validate the effectiveness of the proposed KC-BA in terms of RDP performance, in Fig. \ref{fig_ab_KC_BA}, we provide two variants:
(1) a fixed-length BA strategy without KC-BA, denoted as ``DiT-JSCC w/o KC-BA'', where we manually set $k_s$ and $k_d$ based on the best performing bandwidth proportion results in Fig. \ref{fig_BA_analysis}, allocating the same $k_s$ and $k_d$ to all images; 
(2) an entropy-based BA strategy, denoted as ``DiT-JSCC w/ Entropy-based BA'', we replace the KC score estimation module with a learned entropy model taken from NTSCC \cite{dai2022nonlinear}, i.e., the KC score $\mathcal{I}$ in Eq. \eqref{SC_BA} is substituted by the averaged estimated entropy of the latent features of NTSCC.

From the results in Fig. \ref{fig_ab_KC_BA}, it is evident that compared to the other two variants, KC-BA consistently achieves performance gains across all five consistency and realism metrics, while the entropy-based strategy presents the worst performance.
Intuitively, an effective BA strategy for GenJSCC systems should account for the \textit{conditional generation difficulty} of different image contents. 
As revealed in large-scale generative model analysis \cite{bau2019seeing}, not all semantic classes are equally synthesizable: objects with complex geometric constraints or high intra-class variance are significantly harder for models to generate than texture-dominated objects.
The poor performance of the entropy-based strategy can be attributed to the misalignment between entropy and generative difficulty. 
Standard entropy models tend to allocate excessive bandwidth to high-frequency textures (which have high statistical uncertainty but are easy for DiT to hallucinate) while under-allocating for structural semantics.
In contrast, KC-BA better approximates the inherent semantic complexity of the image, free from the interference of high-frequency details.
This misalignment supports our framework's core principle of prioritizing semantic information during transmission for GenJSCC systems.
Although our method cannot directly estimate the semantic information value, we believe it provides a valuable and practical direction for the future development of GenJSCC systems.

\section{Conclusion and Forward-Looking Discussion} \label{section_conclusion}

This paper has proposed DiT-JSCC, a novel and open-source GenJSCC framework designed to tackle the challenges of semantic inconsistency in image transmission under extreme wireless conditions. By designing a VFM-driven dual-branch encoding structure and customizing a coarse-to-fine CDiT as the generative decoder, DiT-JSCC effectively bridges the gap between semantic representation and generative decoding under image JSCC tasks. Paired with the proposed KC-inspired bandwidth allocation strategy, DiT-JSCC has significantly outperformed existing JSCC methods in perceptual consistency, semantic consistency, and realism, offering a promising direction for robust and high-quality semantic communication.

\add{In future work, we aim to further enhance the proposed framework along several directions. First, we will investigate whether employing more advanced VFMs, such as the latest DINOv3 \cite{simeoni2025dinov3}, can bring additional performance gains. Second, we will evaluate DiT-JSCC on higher-quality and higher-resolution image sources. The current experimental scope is mainly constrained by the available pretrained diffusion models, whose training datasets and native image resolutions affect the datasets and resolution-aligned comparisons that can be evaluated reliably. Extending the framework to broader datasets with more complex scenes and domain shifts will help assess its robustness in open-world semantic communication scenarios. Third, we will study deployment-oriented variants, such as lighter semantic encoders, accelerated diffusion sampling, and model compression, to reduce the computational cost introduced by foundation models and DiT decoding. Finally, perceptual or adversarial objectives, including LPIPS-based feature losses, may further improve local sharpness and realism, but they should be introduced carefully to avoid sacrificing semantic consistency or training stability.}

\ifCLASSOPTIONcaptionsoff
\newpage
\fi

\bibliographystyle{IEEEtran}
\bibliography{Ref}
\end{document}